%% file: tc_extended.tex
\documentclass{article}

\pdfoutput=1 

\usepackage[final,nonatbib]{nips_2017}

\usepackage{tc}

\usepackage{booktabs} 
\usepackage{amsmath}
\usepackage{amsfonts}
\usepackage{nicefrac}
\usepackage{microtype}
\usepackage{xspace,url}
\usepackage{etoolbox} 
\usepackage{multirow}

\usepackage{array}
\newcolumntype{L}[1]{>{\raggedright\let\newline\\\arraybackslash\hspace{0pt}}m{#1}}
\newcolumntype{C}[1]{>{\centering\let\newline\\\arraybackslash\hspace{0pt}}m{#1}}
\newcolumntype{R}[1]{>{\raggedleft\let\newline\\\arraybackslash\hspace{0pt}}m{#1}}

\newcommand{\ic}[1]{\lstinline[language=tc]{#1}}

\def\topsep{\skip0}

\def\labelsep{\skip0}

\newcommand{\figref}[1]         {Figure~\ref{fig:#1}}

\usepackage[binary-units]{siunitx}
\newenvironment{tab}[2][\linewidth]
{\begin{tabular*}{#1}[t]{@{\extracolsep{\fill}}>{\hspace{4pt}}#2}}%
{\end{tabular*}}

\newif\iffinal
\finaltrue 

\definecolor{DarkGreen}{HTML}{33AA33}
\definecolor{OldMauve}{HTML}{673147}
\definecolor{CrimsonRed}{HTML}{990000}
\definecolor{Plum}{HTML}{8E4585}
\definecolor{Sienna}{HTML}{882D17}
\definecolor{Indigo}{HTML}{330088}

\iffinal
\newcommand{\wmnote}[1]{}
\newcommand{\acnote}[1]{}
\newcommand{\aznote}[1]{}
\newcommand{\ntvnote}[1]{}
\newcommand{\ttnote}[1]{}
\newcommand{\zd}[1]{}
\newcommand{\aanote}[1]{}
\else
\newcommand{\wmnote}[1]{{\scriptsize \color{red} [[ Billy: #1 ]]}}
\newcommand{\acnote}[1]{{\scriptsize \color{purple} [[ Albert: #1 ]]}}
\newcommand{\aznote}[1]{{\scriptsize \color{blue} [[ Alex: #1 ]]}}
\newcommand{\ntvnote}[1]{{\scriptsize \color{DarkGreen} [[ Nico: #1 ]]}}
\newcommand{\ttnote}[1]{{\scriptsize \color{CrimsonRed} [[ Theo: #1 ]]}}
\newcommand{\zd}[1]{{\scriptsize \color{Sienna} [[ Zach: #1 ]]}}
\newcommand{\aanote}[1]{{\scriptsize \color{Indigo} [[ Andrew: #1 ]]}}

\fi

\newcommand{\ourtoolkitname}{TC\xspace} 
\newcommand{\isl}{\textit{isl}\xspace}
\newcommand{\ppcg}{\mbox{PPCG}\xspace}
\newcommand{\rstreamtf}{\mbox{R-Stream$\cdot$TF}\xspace}
\newcommand{\pencil}{\textsc{Pencil}\xspace}

\newcommand{\Ocal}{\mathcal{O}}

\usepackage{tikz}
\usetikzlibrary{decorations.pathreplacing,shapes,arrows,positioning,calc}
\newfont{\subheadfnt}{ptmbi8t at 10pt}

\newenvironment{normalitemize}{\setlength{\labelsep}{0.5em}\begin{itemize}}
                              {\end{itemize}}
\newenvironment{normalenumerate}{\setlength{\labelsep}{0.5em}\begin{enumerate}}
                                {\end{enumerate}}

\tikzset{
  pblock/.style={
    rectangle,
    rounded corners,
    draw=black, thin,
    minimum width=2.2cm,
    inner sep=2pt,
    align=left,
    anchor=west,
    fill=yellow,
    font={\fontsize{8.5}{10.2}\selectfont},
    text=black,
    draw=gray,
    thick,
    auto,
    node distance=.2cm and .1cm,
  },
  atune/.style={
    ellipse,
    rounded corners,
    draw=black, thin,
    minimum width=2.2cm,
    inner sep=2pt,
    align=left,
    anchor=west,
    fill=green,
    font={\fontsize{8.5}{10.2}\selectfont},
    text=black,
    draw=gray,
    thick,
    auto,
    node distance=.2cm and .1cm,
  },
  cfedge/.style={
    font=\itshape,
    draw=black,
    ->,
    >=stealth'
  },
  process/.style={
    draw,
    fill=orange!50,
    rectangle,
    minimum height=1.5em,
    minimum width=5em,
    align=center,
    font=\small,
  },
  form/.style={
    draw,
    fill=blue!30,
    ellipse,
    minimum height=1em,
    align=center,
    font=\footnotesize,
  },
  slave/.style={
    execute at end picture={
      \coordinate (lower right) at (current bounding box.south east);
      \coordinate (upper left) at (current bounding box.north west);
      \pgfresetboundingbox
      \path (upper left) rectangle (lower right);
    }
  }
}

\title{Tensor Comprehensions: Framework-Agnostic High-Performance Machine Learning Abstractions}

\author{Nicolas Vasilache\\
  \footnotesize Facebook AI Research\\
  \footnotesize \url{ntv@fb.com}
  \And
  Oleksandr Zinenko\\
  \footnotesize Inria \& ENS, DI\\
  \footnotesize \url{oleksandr.zinenko@inria.fr}\\
  \And
  Theodoros Theodoridis\\
  \footnotesize ETH Z\"{u}rich\\
  \footnotesize \url{theodort@student.ethz.ch}\\
  \AND
  Priya Goyal\\
  \footnotesize Facebook AI Research\\
  \footnotesize \url{prigoyal@fb.com}\\
  \And
  Zachary DeVito\\
  \footnotesize Facebook AI Research\\
  \footnotesize \url{zdevito@fb.com}\\
  \And
  William S. Moses\\
  \footnotesize MIT CSAIL\\
  \footnotesize \url{wmoses@mit.edu}\\
  \AND
  Sven Verdoolaege\\
  \footnotesize \hskip-1cm Polly Labs \& Facebook AI Research\hskip-1cm\,\\
  \footnotesize \url{sven.verdoolaege@gmail.com}\\
  \And
  Andrew Adams\\
  \footnotesize Facebook AI Research\\
  \footnotesize \url{andrew.b.adams@gmail.com}\\
  \And
  Albert Cohen\\
  \footnotesize \hskip-1cm Inria \& ENS, DI \& Facebook AI Research\hskip-1cm\,\\
  \footnotesize \url{albert.cohen@inria.fr}
}

\begin{document}

\maketitle

\begin{abstract}
Deep learning models with convolutional and recurrent networks are now
ubiquitous and analyze massive amounts of audio, image, video, text and graph
data, with applications in automatic translation, speech-to-text, scene
understanding, ranking user preferences, ad placement, etc. Competing
frameworks for building these networks such as TensorFlow, Chainer, CNTK,
Torch/PyTorch, Caffe1/2, MXNet and Theano, explore different tradeoffs between
usability and expressiveness, research or production orientation and supported
hardware. They operate on a DAG of computational operators, wrapping
high-performance libraries such as CUDNN (for NVIDIA GPUs) or NNPACK (for
various CPUs), and automate memory allocation, synchronization,
distribution. Custom operators are needed where the computation does not fit
existing high-performance library calls, usually at a high engineering
cost. This is frequently required when new operators are invented by
researchers: such operators suffer a severe performance penalty, which limits
the pace of innovation. Furthermore, even if there is an existing runtime call
these frameworks can use, it often does not offer optimal performance for a
user's particular network architecture and dataset, missing optimizations
between operators as well as optimizations that can be done knowing the size
and shape of data.  Our contributions include (1) a language close to the
mathematics of deep learning called \emph{Tensor Comprehensions},
(2) a polyhedral Just-In-Time compiler to
convert a mathematical description of a deep learning DAG into a CUDA kernel
with delegated memory management and synchronization, also providing
optimizations such as operator fusion and specialization for specific sizes,
(3) a compilation cache populated by an autotuner. In particular, we
demonstrate the suitability of the polyhedral framework to construct a
domain-specific optimizer effective on state-of-the-art deep learning models
on GPUs. Our flow reaches up to $4\times$ speedup over NVIDIA libraries on
kernels relevant to the Machine Learning Community, and on an actual model used
in production at Facebook.
It is integrated with mainstream frameworks Caffe2 (production-oriented),
PyTorch (research-oriented), through the ATen asynchronous tensor library.
\end{abstract}

\input{1-introduction}

\input{8-related-and-future-work}

\input{3-tensor-comprehensions}

\input{2-workflow}

\input{4-polyhedral-compiler}

\input{5-tuning-and-caching}

\input{7-examples}

\input{9-conclusion}

\bibliography{pldi2018}

\input{a-appendix}

\end{document}

%% file: 1-introduction.tex
\section{Introduction}

Deep neural networks trained with back-propagation learning~\cite{Backprop89}
are a method of choice to solve complex problems with sufficient
data. Recently, GPU-accelerated algorithms have excelled in this
area~\cite{Raina:2009:LDU:1553374.1553486,DBLP:journals/corr/abs-1003-0358,Alexnet12}.
Popular computation graph engines \cite{Theano,Torch7,MXNet,TensorFlow} offer
high-level abstractions for optimizing and executing deep neural networks
expressed as graphs of tensor operations. These frameworks make transparent
use of GPUs and other hardware accelerators for low power or low latency
\cite{Brainwave17,TPU17} and are often implemented as an abstraction over
highly-optimized routines for individual operators.
While these operators are sufficient for many applications, they fall short in
a number of instances where the computation does not fit the supported library
calls. Consider a researcher who wants to develop a novel type of layer or
network architecture. She must develop a custom operator, often at a high
engineering cost and performance penalty. Furthermore, even when it is
possible to represent a given network with library calls, they often miss peak
performance for two reasons: missed optimizations across operators, and no
tuning for every combination of size, shape and data flow encountered in
Machine Learning (ML) \cite{FBFFT15}.

Alone, computation graphs in such frameworks are too abstract to capture
essential refinements and lowering steps required for efficient use of
hardware accelerators, unless the operators perfectly fit a pre-optimized set
of library functions. The parallel execution of individual layers and the
memory layout of individual tensors varies greatly depending on data size and
shape, upstream and downstream computations, and specific hardware features.

\subsection{Motivations}
These observations have pushed for an active library
\cite{VG98,Bec03} or built-to-order (BTO) approach \cite{BTO09}, in which library code
is specialized and generated on-demand.
However, this approach does not quite solve the problem as tuning library
kernels in isolation misses context-dependent opportunities and creating a
library that covers all combinations of individual kernels is infeasible.

This has led to the creation of domain-specific languages such as
Halide~\cite{Halide}, which has been successful in imaging due to its
ability to fuse large pipelines without obfuscating the underlying
algorithm. However when using Halide on the GPU, all scheduling
transformations must be manually specified, and achieving high
performance with the right combination of them is beyond the ability
of most users.

More recent deep learning compilers such as XLA~\cite{XLA} and
Latte~\cite{Latte} seem to be the ideal solution to this problem: they
combine operators from computation graphs, allowing for optimizations
across operators as well as optimizations that take advantage of data
size. Yet, so far, the expected performance levels have not been met
on GPU targets.  The transformation language of these frameworks does
not seem to be able to represent complex scheduling and mapping
transformations which are often crucial to GPU targets with
partitioned memory architectures.

To remedy this, an effective programming language for computation graph
engines must simultaneously address the two following challenges:
\begin{normalitemize}
\item ensure that abstraction not only enhances programmer
  productivity but also enables the compiler and its supporting execution
  environment to eliminate concerns irrelevant to the target platform, to refine
  the code through intermediate representations closer to the machine, and to
  automatically explore a wide optimization space. In other words, the system must be able to
  offer ``abstraction without regret''
  \cite{Rompf:2010:LMS:1868294.1868314,Coh06} while conveying rich semantical
  information available at
  compilation time;
\item select appropriate intermediate representations and optimization
  algorithms that deal with deep parallelism and memory hierarchies, as well
  as hardware features such as vector instructions and special-purpose memory.
\end{normalitemize}

\subsection{Contributions}
We present a novel domain-specific flow capable of generating highly-optimized
kernels for tensor expressions, leveraging optimizations across operators and
optimizations that take into account the size and shape of data.
We address the first challenge through the design of Tensor Comprehensions~%
(TC), a domain-specific language whose syntax is both concise and expressive
and whose semantics allows for efficient memory management and mapping to
complex parallel platforms. We address the second challenge by specializing a
polyhedral intermediate representation and its compilation algorithms to the
domain of deep learning, providing it with a dedicated autotuner.
The polyhedral framework of compilation emerged as a natural candidate to
design a versatile intermediate representation and optimization flow
satisfying the needs of the domain and target hardware. The polyhedral
framework has demonstrated strong results in domain-specific optimization
\cite{Polymage,VOBLA,Baghdadi2015Pencil}, expert-driven
metaprogramming~\cite{URUK,CHiLL,Clay}, libraries of high-level
transformations of control flow and storage \cite{loopy},
and embedding of third-party library code
\cite{DBLP:conf/pldi/KongVSFPS13}, and automatic generation of efficient code
for heterogeneous targets
\cite{PlutoGPU,RStream,PouchetFPGA,PPCG2013,Baghdadi2015Pencil,RR-9110}.
\begin{tclisting}
\end{tclisting}
In this report, we present the following contributions:
\begin{normalitemize}
\item Tensor Comprehensions (TC): a high-level language to express tensor
  computations arising in ML with a syntax generalizing the Einstein
  notation. It supports shape and size inference, flexible element-wise syntax
  with both named and positional parameters. It has
  conciseness and safety advantages, avoiding off-by-one errors while also
  allowing layout transformations and specialization.
\item An end-to-end compilation flow capable of lowering tensor comprehensions
  to efficient GPU code. It delivers strong baseline performance for custom
  operators and remains competitive with vendor libraries on standard
  ones. The former is essential to reducing the technical debt on vendor
  libraries, enabling ML researchers to explore a wider field of architectures
  and layers in production-like scenarios.
\item A collection of polyhedral compilation algorithms with a specific domain
  and target orientation. Unlike general-purpose parallelizing compilers, we
  primarily optimize for reduced launch and synchronization overhead through
  kernel fusion and also favor multi-level parallelism and promotion to deeper
  levels of the memory hierarchy.
\item An autotuning framework that takes advantage of Just-In-Time (JIT)
  compilation and code caching. It includes specialization for non-standard
  sizes, eliminating control and address generation logic, and takes ownership
  of all optimization knobs from the ML framework to the code generator.
\item Integration into two common ML frameworks (PyTorch \cite{PyTorch} and
  Caffe2 \cite{Caffe2}). In principle our system is general enough to be
  integrated into other ML frameworks.
\end{normalitemize}
For our initial system, we focus on the generation of CUDA code because NVIDIA
GPUs dominate the hardware landscape for training deep neural networks. We
believe our approach applies to other types of heterogeneous nodes with shared
or partitioned memory.

The report comes with supplementary material labeled as
Section~\ref{sec:appendix}, covering background on polyhedral compilation and
deep learning frameworks, implementation details and further experimental
methodology.\footnote{The report is meant to be easily accessible to a reader
  familiar with parallelizing compilation, performance tuning for GPUs, and a
  basic knowledge of the polyhedral framework, tensor algebra and convolution
  operations.}

%% file: 8-related-and-future-work.tex
\section{Related Work}

Despite decades of progress in optimizing and parallelizing
compilation, programmers of computationally intensive applications
complain about the poor performance of optimizing compilers, often
missing the peak achievable performance by orders of magnitude. Among
the reasons for this state of affairs, one may cite the complexity
and dynamic behavior of modern processors, domain knowledge required
to prove optimizations' validity or profitability being unavailable to
the compiler, program transformations whose profitability is difficult
to assess, and the intrinsic difficulty of composing complex
transformations, particularly in the case of computationally intensive
loop nests \cite{Coh05,Don05,URUK,CHiLL,loopy,Clay}.

Several contributions have successfully addressed this issue, not by improving
a general-purpose compiler, but through the design of
application-specific program generators, a.k.a.\ active libraries
\cite{VG98}. Such generators often rely on feedback-directed
optimization to select the best generation strategy \cite{Smi00}, as
popularized by ATLAS \cite{ATLAS} for dense matrix operations---and more
recently BTO \cite{BTO09}---and FFTW
\cite{FFTW} for the fast Fourier transform.
Most of these generators use transformations previously proposed for
traditional compilers, which fail to apply them for the aforementioned
reasons. The SPIRAL project
\cite{SPIRAL} made a quantum leap over these active libraries, operating on a
domain-specific language (DSL) of digital signal processing formulas.
Compilers for DSLs typically rely on domain-specific constructs to capture the
intrinsic parallelism and locality of the application.  Using such an
approach, DSL compilers such as Halide~\cite{Halide} for image processing show
impressive results. Its inputs are images defined on an infinite range, while
TC sets a fixed size for each dimension using range inference. This is better
suited to ML applications, which mostly compute on fixed size tensors with
higher temporal locality than images; it is also less verbose in the case of
reductions and does not carry the syntactic burden of pre-declaring stage
names and free variables (Halide needs this as a DSL embedded in C++).
OoLaLa~\cite{OoLaLa} takes a similar approach for linear algebra, and
TACO~\cite{Taco} and Simit~\cite{Simit} use a notation similar to that of TC,
but generate sparse matrix code for numerical solvers.

Following this trend in the context of deep neural networks, we not only
design yet another DSL and compiler but propose a more generic code generation
and optimization framework bringing together decades of research in loop nest
optimization and parallelization for high-performance computing. We also
design the domain language to cover a variety of existing and emerging machine
learning models.
Our framework automates a combination of affine transformations involving
hierarchical tiling, mapping, shifting, fusion, distribution, interchange, on
either parametric or fully instantiated problems, that are not accessible to
Halide \cite{Halide,Mullapudi2016HaideAutoscheduler}, Latte \cite{Latte} or
XLA's \cite{XLA} representations of tensor operations.

The polyhedral framework is a powerful abstraction for
the analysis and transformation of loop nests, and a number of tools and
libraries have been developed to realize its benefits
\cite{feautrier92multi,Bondhugula2008Pluto,PPCG2013,Bondhugula2016Pluto+,RR-9110},
including components for
production compilers such as GCC (Graphite) and LLVM (Polly).
Polyhedral techniques have also been tailored for domain-specific
purposes. State of the art examples include the PolyMage~\cite{Polymage} DSL
for image processing pipelines and the PENCIL approach to the construction of
parallelizing and compilers for DSLs~\cite{Baghdadi2015Pencil,VOBLA}.
Interestingly, some optimization techniques derived from PolyMage crossed over
from polyhedral representations into Halide's recent automatic
scheduler~\cite{Mullapudi:2016:ASH:2897824.2925952}.  Our compiler implements
optimizations specific to the long, non-uniform reuse patterns and deeply
nested loops of deep learning models; these heuristics are not available in
Halide and variants \cite{Polymage,Mullapudi2016HaideAutoscheduler}.





Back to deep learning frameworks, TC shares several motivations with Latte
\cite{Latte}, including a high level domain-specific language and an
end-to-end flow.
TC provides an element-wise access that is just as expressive when implementing
custom layers, but unlike Latte it is more concise thanks to type and shape
inference, safer regarding static bound checking and graph connectivity, and
more flexible by decoupling indexing from representation and layout choices
(e.g., sparse layers). In addition, our framework implements more complex
scheduling and mapping transformations than Latte, some of which are essential
to GPU targets with partitioned memory architectures. Unlike Latte, it is also
designed as a JIT compilation library for seamless integration with deep
learning frameworks.

Like tensor comprehensions, XLA \cite{XLA}
provides automatic shape and size inference, it may operate ``in process''
as a JIT compilation library, and it integrates into a production deep
learning framework (TensorFlow, Caffe2~\cite{Caffe2}). XLA shares many
motivations with Latte, with a focus on integration and completeness of
functionality rather than on the complexity of the optimizations and mapping
strategies. Most of our design and algorithmic contributions would naturally
fit XLA, except for the following: \ourtoolkitname remains independent from a
specific computation graph framework while preserving tight integration with
production frameworks; we did not use an embedded DSL approach---keeping C++
as an interface for implementing optimization strategies only---isolating the
user from complexity and debugging hurdles of embedded DSLs.

Most recently, \rstreamtf \cite{RStreamTF} was presented as a proof-of-concept adaptation of the R-Stream polyhedral compiler to the automatic optimization of TensorFlow operators. Similarly to our approach, the generated code is wrapped as a custom operator of TensorFlow. The tool takes a computation graph as input and partitions it into subgraphs amenable to tensor fusion, contraction and layout optimization. \rstreamtf also leverages the broadcast semantics of TensorFlow to maximize the operator's polymorphism w.r.t.\ input tensor dimension and shapes. This makes \rstreamtf very aggressive in terms of static memory management and kernel partitioning.
We made the more pragmatic choice of leaving most of these decisions
to the level of tensor algebra, allowing a domain-specific optimizer or ML expert to rewrite declarative comprehensions into capacity- and layout-optimized ones. On the contrary, \ourtoolkitname is more ambitious in its domain-specialization of affine scheduling and mapping, aiming for the generation of a single accelerated kernel, with heuristics adapted to the high dimensionality and non-uniform, long reuse patterns of neural networks. The lack of algorithmic detail in the paper does not let us compare those affine transformation heuristics at the time of writing.

%% file: 3-tensor-comprehensions.tex
\section{Tensor Comprehensions}
\label{sec:tc}

We provide a language for expressing element-wise computations of
tensors using \emph{tensor comprehensions}.

Tensor Comprehensions (TC) are a notation for computing on multi-dimensional
arrays that borrows from the Einstein notation (a.k.a.\ summation convention):
\begin{normalenumerate}
\item index variables are defined implicitly by using them in an expression
  and their range is inferred from what they index;
\item indices that appear on the right of an expression but not on the left
  are assumed to be reduction dimensions;
\item the evaluation order of points in the iteration space does not affect
  the output.
\end{normalenumerate}
Let us consider matrix-vector product as a simple example of a tensor
comprehension with two statements:
\begin{tclisting}
def mv(float(M,K) A, float(K) x) -> (C) {
  C(i)  = 0
  C(i) += A(i,k) * x(k)
}
\end{tclisting}

This defines the function \ic{mv} with \ic{A} and \ic{x} as input tensors and
\ic{C} as output. The statement introduces two index variables `\ic{i}' and
`\ic{k}'. Variables not defined anywhere, implicitly become index
variables. Their range is inferred based on how they are used in indexing (see
below); here we will discover \ic{i = [0,M)}, and \ic{k = [0,K)}.
Because \ic{k} only appears on the right-hand side, stores into \ic{C} will
\emph{reduce} over \ic{k} with the reduction operator \ic{+}. Reductions can
occur across multiple variables, but they all share the same kind of
associative and commutative operator (e.g., \ic{+=}) to ensure that evaluation
order does not affect the computed value (e.g., composition of
$\min$ and $\max$ do not commute, $\min(\max(f(.))) \neq \max(\min(f(.)))$).

Intuitively, a tensor comprehension may be thought of as
the \emph{body} of a loop
whose control flow is inferred from context. The equivalent C-style
pseudo-code is:
\begin{clisting}
tensor C({M}).zero(); // 0-filled single-dim tensor
parallel for (int i = 0; i < M; i++)
  reduction for (int k = 0; k < K; k++)
    C(i) += A(i,k) * x(k);
\end{clisting}
Importantly, the nesting order (\ic{i} then \ic{k}) is arbitrary: the
semantics of a tensor comprehension is always invariant by loop permutation.

TC allows in-place updates, but preserves a functional semantics that is atomic
on full tensors: the semantics is to \emph{read RHS expressions in full before
  assigning any element on the LHS}. This specification is important in case
the LHS tensor also occurs in RHS \cite{FRAGUELA2012465}: the compiler is
responsible for checking the causality of in-place updates on element-wise
dependences. We currently implement a simple syntactic check, allowing only
in-place updates on pointwise definitions and tensor contractions. When this
check fails, the compiler rejects the program due to liveness interference;
for example, any in-place transposition \ic{a(i,j) = a(j,i)} is incorrect
(unless the range is empty or a single element), while \ic{a(i,j) = b(j,i)} is
a valid transposition. Of course, this explicit reuse and atomic update
semantics does not preclude other scheduling and storage mapping decisions by
the compiler, as long as these preserve the element-wise dependences of the
TC. This mixed declarative-imperative design of TC is inspired from
Lush~\cite{lush2002}.

\begin{figure}
\begin{tclisting}
def sgemm(float a, float b,
          float(N,M) A, float(M,K) B) -> (C) {
  C(i,j)  = b * C(i,j)            # initialization
  C(i,j) += a * A(i,k) * B(k,j)   # accumulation
}
\end{tclisting}
\caption{Tensor Comprehension for the \ic{sgemm} BLAS}
\label{fig:tc_sgemm}
\end{figure}

Reductions in TC are often initialized with the reduction operator's neutral
element; we provide a short-cut for an \emph{initializing reduction},
appending `\ic{!}' to the reduction symbol, i.e., `\ic{+=!}' instead of
`\ic{+=}'. Here is a one line definition of the matrix-vector product:
\begin{tclisting}
def mv(float(M,K) A, float(K) x) -> (C) {
  C(i) +=! A(i,k) * x(k)
}
\end{tclisting}
Using these simple properties, common machine learning kernels can be written
in just a few lines. For instance, Figure~\ref{fig:tc_sgemm} shows the SGEMM
function of the BLAS library.
General tensor contractions can be expressed along the same lines.
A fully connected layer followed by a rectified linear unit takes the form of
a transposed matrix multiplication initialized to a broadcast bias term and
followed by pointwise clamping (i.e., \ic{fmaxf} with $0$):\footnote{For
  historical reasons related to the expression of element-wise linear algebra
  in the context of neural networks, the expression in matrix form often
  involves transpositions. TC lifts that impedance mismatch and makes it
  non-surprising: the tensor layout and sizes must match what is expected from
  the indexing expression.}
\begin{tclisting}
def fcrelu(float(B,I) in, float(O,I) weight,
           float(I) bias) -> (out) {
  out(i,j)  = bias(j)
  out(b,o) += in(b,i) * weight(o,i)
  out(i,j)  = fmaxf(out(i,j), 0)
}
\end{tclisting}
Here we chose to reuse the \ic{out} tensor across all comprehensions, indicating the absence of temporary storage.

A 2-D convolution is similarly simple, its reduction is initialized to $0$ (note the use of \ic{+=!}):
\begin{tclisting}
def conv2d(float(B,IP,H,W) in,
           float(OP,IP,KH,KW) weight) -> (out) {
  out(b,op,h,w) +=! in(b,ip, h + kh, w + kw)
                    * weight(op,ip,kh,kw)
}
\end{tclisting}
with reduction dimensions \ic{kh}, \ic{kw}. A max pooling layer is:
\begin{tclisting}
def maxpool2x2(float(B,C,H,W) in) -> (out) {
  out(b,c,i,j) max=! in(b,c, 2 * i + kw, 2 * j + kh)
    where kw in 0:2, kh in 0:2
}
\end{tclisting}
In the case of max pooling, the indexes \ic{kw} and \ic{kh}, which determine
how many entries to pool over, are under-constrained since they are not
inferable from any input tensors and the range inference procedure emits an
error when no further information is provided about these indices.
So we provide a \ic{where} annotation to inform the inference algorithm
of the intended ranges of these variables and let it infer the remaining
ranges from context.

Subscript expressions can be any affine function of iterators, or subscript-of-subscript expressions, and combinations thereof. The latter capture data-dependent accesses such as a gather operation:
\begin{tclisting}
def gather(float(N) X, int(A,B) I) -> (Z) {
  Z(i,j) = X(I(i,j))
}
\end{tclisting}

TC closely matches an algorithmic notation. This is not true of today's
prominent frameworks where most operators are defined as black-box
functions. The design of TC makes it easy to experiment with small layer
variations, while preserving a concise, in-place expression. For instance,
recently, strided convolutions have become popular in image
classification~\cite{ResNext}. With tensor comprehensions, a strided
convolution is easily created as a tweak on convolution, here is a convolution
strided by \ic{sh} along \ic{h} and \ic{sw} along \ic{w}:
\begin{tclisting}
def sconv2d(int sh, int sw, float(N,C,H,W) I,
            float(F,C,KH,KW) W, float(F) B) -> (O) {
  O(n,f,h,w)  = B(f)
  O(n,f,h,w) += I(n,c, sh * h + kh, sw * w + kw)
                * W(f,c,kh,kw)
}
\end{tclisting}
Note that efficient implementations of \ic{sconv2d} can take advantage
of partial evaluation for the frequent case where \ic{sh} and \ic{sw} are
constant, resulting in affine subscript expressions. This is another advantage
of JIT compiling TC.

\subsection{Range Inference}
Tensor comprehensions are concise because most of the time loop ranges are
inferred from context. Similar to the inference of polymorphic data types, the
inference algorithm aims to accurately infer the range based on usage
patterns. It is not always possible
however, due to non-affine expressions or under-constrained problems---see
e.g., max pooling. TC needs additional annotations for such cases.

Iteration variables are always non-negative. Unless specified otherwise in a
\ic{where} clause, each iteration variable is assumed to start at 0.

Inference is computed from input arguments to output tensors, setting up a
constraint-based analysis problem across all the affine array accesses in a TC
function. We initially looked for a very general setting of this problem, with
an inference algorithm relying on Presburger arithmetic as implemented in
polyhedral libraries \cite{Verdoolaege2011iscc}.  Yet, to program
productively, one must be able to mentally emulate the written code on the
abstract machine defined by the language semantics. If this requires more
thought than writing explicit loops would, our language design has
failed. With this in mind, we eschew heavy-duty mathematical tools, and take a
more straightforward approach for the sake of usability. We infer ranges only
in cases where we feel they are obvious, and require explicit annotation
elsewhere. We intend to fine-tune this boundary in the future depending on
what users find surprising.

To find a good first approximation,
we infer rectangular ranges that are as large
as possible without reading out of bounds on the inputs. If there is more than
one way to do this, we throw an error and require explicit annotation using a
'where' clause.

This rule is sufficient to understand the \ic{sgemm} case above. Maximizing
the range of \ic{i} gives it the range of the number of rows of
\ic{A}. Similarly maximizing the range of \ic{j} gives it the range of the
columns of \ic{B}. \ic{k} is used twice, so making \ic{k} as large as possible
gives it the lesser of the number of columns of \ic{A} and the number of rows
of \ic{B}. These in turn are constrained to be equal by the type signature of
the function (they are both \ic{K}).

Now consider a simple 1-D convolution:
\begin{tclisting}
def conv1d(float(M) I, float(N) K) -> (O) {
  O(i) = K(x) * I(i + x)
}
\end{tclisting}

There are multiple ways in which we could maximize the ranges of \ic{i}
and \ic{x}. If we first maximize \ic{i}, we might say that it ranges over
the entirety of \ic{I}. This forces \ic{x} to take on a single value only,
which will not result in the output one expects from a convolution, as it
ignores most of the kernel! If we first maximize \ic{x}, so that it
ranges over the entirety of K, then in order to not read out of bounds
the range of \ic{i} must be smaller, and we get an output that is
slightly smaller than the input. This is the behavior we prefer.

In order to make this unambiguous without requiring explicit
annotation in this simple case, range inference proceeds in rounds. We
maintain a set of unresolved variables. Initially, this set contains all
variables not constrained with an explicit \ic{where} clause. In each
round, we consider all the tensor argument expressions that contain a
single unresolved variable, and construct a boolean expression that
states the access is not out-of-bounds. We then use tools from
Halide (\ic{solve_for_inner_interval}) to find the maximal range for
the variable that satisfies this condition, given the ranges of
variables already resolved in previous rounds. If the variable was
already constrained in this round by some other tensor access, we take
the intersection of the inferred ranges.

For the stencil above, in the first round we ignore the expression \ic{I(i+x)}
because it contains multiple unresolved variables. We use the expression
\ic{K(x)} to deduce a range for \ic{x}. In the second round, \ic{I(i+x)} now
contains a single unresolved variable, and we use the already-inferred range
of \ic{x} to deduce a maximal range for \ic{i}. In other words, the access
\ic{K(x)} constrains \ic{x} to range from 0 to $N-1$, while the access
\ic{I(i+x)} constrains \ic{i+x} to range from 0 to $M-1$, from which we deduce
that \ic{i} must satisfy the following constraint
$$\forall x \in [0, N-1], i+x\in [0, M-1],$$
which yields the final range of 0 to $M-N$ for \ic{i}.

Notice the $\forall$ universal quantification on $i$, constraining the output
tensor \ic{O} to be uniformly defined across the entire domain from the exact
same inputs in \ic{I}. This differs from the approach taken in the Alpha
language for systems of affine recurrent equations \cite{Alpha} where the
domain scanned by $x$ would be the projection of the triangular domain
of $(i, x)$ involving existential quantification:
$\{x \mid x\geq 0 \wedge \exists i \in [0, N-1], i+x\in [0, M-1]\} = [0, M-1]$.
The semantical choice of quantifying iterators universally is derived from
Halide; it guarantees that the shape of input tensors is always a
(hyper-)parallelepiped, it is intuitive to ML experts and facilitates the
generation of compact code, free of conditional control flow---unlike a more
conventional projection semantics.

More complex examples involve intersecting successive rounds of inference over
the same variables, and also ambiguous cases where no single rectangular shape
can be derived from a more general set of constraints. The latter will not
type-check and the compiler will request the insertion of a disambiguating
\ic{where} clause.

In addition, since memory management is externalized to a ML framework, the
list of temporary tensors defined within a TC must be made explicit---with
their shape and ranges; the range inference algorithm infers this list
automatically. Finally, the inference algorithm can be formalized as a type
and effect system \cite{NNH99}, collecting constraints on index sets along the
sequence of tensor definitions.

In its first release, TC does not support recurrent definitions such as those
needed for the implementation of Recurrent Neural Networks (RNN).

Note that tensor indexing may involve any integer-valued expression, as long as it does not depend on the LHS tensor of the current statement. The compiler does its best at range inference and in the static verification that individual elements are defined only once in an imperatively-defined dimension; compilation fails in case range inference does not succeed and no \ic{where} clause is there to provide the missing information.

\subsection{Data Layout Transformations}
TC makes global data layout transformations significantly easier. The ML
community has been heavily relying on such transformations by composing
operations on tensor metadata in the form of a tuple of $(\mbox{dataPtr},
\mbox{offset}, \mbox{size}[], \mbox{stride}[])$ \cite{BTO09}. One of the main
usage ML researchers have for such primitives is algorithmic tiling and
hierarchical decompositions. The former has strong connections to data tiling
and is now ubiquitous in the implementation of high-performance convolutions
in the frequency domain to keep the memory footprint under
control~\cite{FBFFT15} or to fit to fast local memories~\cite{Winograd}. The
\emph{unfold} operation in Lush~\cite{lush2002} and Torch~\cite{Torch7}
perfectly matches implicit data tiling (i.e., without explicit memory copies),
when there are no partial border effects. TC supports generalized tensor
transpositions (i.e., applying an $n$-D permutation matrix where $n>2$) and
data tiling can be achieved by simply reshaping tensors and adjusting the TC
index expressions.
Range inference and checking guarantees such reshaping will always be
consistent throughout a TC.
Array-of-struct to struct-of-array conversions---and similar
operations on short vectors---are available by-products: they
are particular cases of dimension interchange and data tiling. At this
time, data layout and TC transformations are left to the domain expert
at source level, the TC inference procedure guarantees the expressions
have the proper ranges. We will reevaluate this assumption in the
future as we introduce automatic data layout transformations in
\ourtoolkitname.

%% file: 2-workflow.tex
\section{High-Level Workflow Description}
Let us position our work in the context of deep learning frameworks such as
TensorFlow, Caffe2 and PyTorch.\footnote{See Section~\ref{sec:mlframeworks}
  for a brief introduction to these frameworks.}

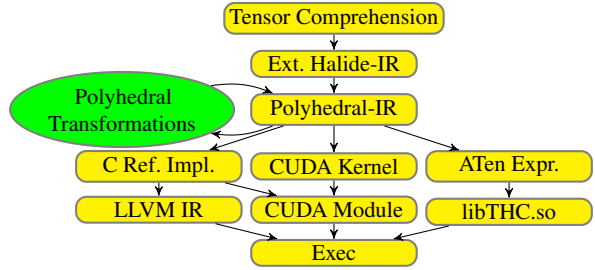
\begin{figure}[t]
\begin{minipage}[T]{\linewidth}
  \centering
  \scalebox{1}{
        \begin{tikzpicture}[slave]
          \node [pblock] (TC) {Tensor Comprehension};
          \node [pblock, below=of TC] (Halide) {Ext.\ Halide-IR};
          \node [pblock, below= of Halide] (Poly) {Polyhedral-IR};
          \node [atune, minimum width=1cm, left= of Poly, xshift=-0.1cm] (PolyTransfos) {\hskip10pt Polyhedral\\ Transformations};
          \node [pblock, below= of Poly, yshift=-.15cm] (CUDA) {CUDA Kernel};
          \node [pblock, left= of CUDA] (C) {C Ref.\ Impl.};
          \node [pblock, below=of CUDA] (Module) {CUDA Module};
          \node [pblock, left=of Module] (LLVM) {LLVM IR};
          \node [pblock, right= of CUDA] (ATen) {ATen Expr.};
          \node [pblock, below=of ATen] (THC) {libTHC.so};
          \node [pblock, below=of Module] (Exec) {Exec};
          {D};
          \path [cfedge]
          (TC)  edge node {} (Halide)
          (Halide)  edge node {} (Poly)
          (Poly)  edge[bend left=15] node {} (PolyTransfos)
          (PolyTransfos)  edge[bend left=15] node {} (Poly)
          (Poly)  edge node {} (C)
          (Poly)  edge node {} (CUDA)
          (Poly)  edge node {} (ATen)
          (C)  edge node {} (LLVM)
          (C)  edge node {} (Module)
          (CUDA)  edge node {} (Module)
          (ATen)  edge node {} (THC)
          (LLVM)  edge node {} (Exec)
          (Module)  edge node {} (Exec)
          (THC)  edge node {} (Exec);
        \end{tikzpicture}
}
    \end{minipage}
  \caption{The JIT compilation flow lowers TC to Halide-IR, then to Polyhedral-IR, followed by optimization, code generation and execution}
  \label{fig:flow}
\end{figure}

TC expressions are first integrated into the ML framework as follows. We opted
for an ``in process'' implementation, streamlining the interaction with
computation graph engines and ML applications built on top of them, a unique
feature for a fully-automated scheduling and mapping flow. We provide a thin
API that translates the specific tensor object model to our own. Operator
definitions are overridden to generate TC rather than the framework's backend
implementation, as well as provide users the ability to write their own
TC.\footnote{Section~\ref{sec:interface-api} describes the programming
  interface for TC.}
In this context, a single TC may correspond to a DAG of operators in the ML
framework.  This matching is currently done manually. Automatic DAG
partitioning, matching and rewrites (like e.g., TensorRT~\cite{TensorRT}),
informed by our compilation flow, are left for future work. The TCs are then
JIT compiled as shown in \figref{flow}. In cases where the default backend
implementation performs better, we fall back to the reference implementation.

Starting from a TC with specialized tensor sizes and strides,\footnote{Our
  toolchain supports parametric specifications, yet we have found early
  specialization to be beneficial in driving profitability decisions during
  polyhedral scheduling.} we lower it to a parametric Halide expression.

In a prototype version of the system, Halide-IR was lowered to polyhedral
representation via a translation to PENCIL~\cite{Baghdadi2015Pencil} and then
parsed using the pet~\cite{PET} and isl~\cite{ISL10} libraries. To reduce the
impedance mismatch between IRs and to facilitate the propagation of
semantical annotations, the flow evolved to lower Halide-IR directly
to a polyhedral representation, bypassing the PENCIL intermediate language.

Similarly, we bootstrapped CUDA kernel generation in the original prototype
from a modified version of the \ppcg\ compiler~\cite{PPCG2013}. The
modifications included its usage as a library, for in-process, multi-threaded
operation (see Section~\ref{sec:polyhedral}). When moving away from PENCIL,
the functions of \ppcg\ were reimplemented from the ground up, aiming for more
modularity and compliance with a modern C++ environment.

Complementing this flow, an autotuner and a serializable compilation tightly
interact with scheduling and mapping strategies to search the transformation
space (see Section~\ref{sec:autotuning}).


Additionally, we provide a simple ``identity'' polyhedral mapping option to
generate a naive, readable, CUDA reference implementation that may be run and
checked for correctness on a single GPU thread and shake off simple
problems.
We are planning a simple LLVM JIT compilation pass to make that reference
implementation more generally usable on a CPU, but this is not yet implemented.

Lastly, we will provide a path to emit a series of library calls, a useful
fallback to default implementations backed by CUDNN.

%% file: 4-polyhedral-compiler.tex
\section{Polyhedral JIT Compilation\label{sec:polyhedral}}
%
%
%
The compiler bridges the impedance mismatch between the logical layout of high
level tensor operations (dimension ordering) and the data format the
polyhedral code generator expects (C99 arrays~\cite{Baghdadi2015Pencil}). The
lowering step ensures combinations of size and stride correspond to
non-aliasing array and subarray syntax;
it also ensures the absence of out-of-bounds access, analyzing access relations and inferred tensor ranges;
it emits tensor declarations and reorder expressions to match the data model of the target language, i.e., row-major arrays.

Note that tensor specifications may feature input and output arguments
aliasing for in-place computation and implicit conversion of tensors of higher
dimension. We argue that such specifications should lead to multi-versioning
and specialization for each aliasing scenario. Also, the semantics adopted
by TC, building on range inference, differs from NumPy-style
``broadcast'' semantics adopted in some form or
another by XLA, PyTorch and MXNet.\footnote{Broadcasting is a set of
  non-trivial rules that allow implicit conversion between tensors of
  different dimensions. It enables certain tensor operations even when an
  appropriate library implementation does not exist for those non-conforming
  shapes. It carries its baggage and ambiguities when dealing with higher
  dimensional tensor contractions, as demonstrated in the TensorFlow Github
  issue \#5523.} TC does not need such implicit syntactic sugar. For example,
the TC corresponding to the so-called outer product matrix multiplication \ic{[p,q,r] matmul [1,s,r,t] -> [p,s,q,t]} is simply:
 \begin{tclisting}
def outerProductMM(float(P,Q,R) A, float(S,R,T) B) -> (O) {
  O(p,s,q,t) +=! A(p,q,r) * B(s,r,t)
}
\end{tclisting}
One may further transform layouts and derive a \ic{QPTS} version (named by the ordering of output dimensions) instead of \ic{PSQT} if desirable.

Additional lowering steps include forward substitution of convolution expressions (storage/computation tradeoff), padding with zero, mirroring and clipping.

The \emph{polyhedral framework} offers a state-of-the-art internal compiler
representation of loop-based programs that are ``mostly
regular''~\cite{Feautrier2011Polyhedron}.
In its most basic form, it describes arithmetic expressions surrounded by
loops and branches whose control flow can be determined statically. Hence the
polyhedral framework is said to operate on \emph{static control parts} (SCoP)
of the program.
More specifically, loop bounds and branch conditions are affine expressions of
outer loop iterators and statically unknown constant values that are treated
symbolically and referred to as \emph{parameters}.
Computation is described using arithmetic expressions on array elements with
the same restrictions on subscripts as on loop bounds.
Extensions exist to handle irregularities through
over-approximation~\cite{Benabderrahmane2010Polyhedral} or user-defined
annotations~\cite{Baghdadi2015Pencil}.
Despite its deceiving apparent simplicity, the framework captures large
classes of computation-intensive codes, it is parametric on domain and array
sizes, and more expressive than domain-specific representations such as Halide's.

\emph{This work demonstrates that the polyhedral framework is particularly
  well suited for deep neural networks, associated with large and deeply
  nested loops with long dependence chains and non-uniform or all-to-all
  patterns---arising from fully connected layers and tensor contractions and
  transpositions. These features push the optimization problem into a
  different heuristic space than Halide's for image processing, and a much
  wider space than linear algebra alone.}

We use the \emph{named relation notation} introduced in
\ic{iscc}~\cite{Verdoolaege2011iscc} for unions of relations where tuples of
iterators are guarded with syntactic identifiers~\cite{Pugh1994Static}.
 The reader unfamiliar with polyhedral compilation---iteration domains, affine
 access and dependence relations, scheduling and polyhedral code
 generation---may refer to Section~\ref{sec:polyhedral_background} in the
 supplementary material.

\paragraph{Schedule Trees}
Affine maps can be composed into \emph{schedule trees}
\cite{Verdoolaege2014ScheduleTrees} to facilitate the communication of
properties from the high-level language (here, TC) to the downstream optimizer
and to attach target-specific information along the way
(e.g., SPMD thread-relative induction as in CUDA, synchronization,
data transfer instructions~\cite{PPCG2013,Baghdadi2015Pencil}).
Schedule trees introduce specific node types.
A \emph{band node} defines a \emph{partial} execution order through one or
multiple piecewise affine functions defined over iteration domains.
The name refers to the notion of a \emph{permutable schedule band}, a tuple of
one-dimensional schedule functions that can be freely interchanged while
preserving the semantics of the program.
An affine function in a band is often referred to as
a \emph{band member} or a \emph{schedule dimension}.
A collection of \emph{filter nodes}
partitions the iteration space, binding its subtree to a
subset of the iteration domain. They can be arranged into \emph{set or sequence
  nodes} depending on whether the order of execution must be constrained for
correctness or not (i.e., whether or not it corresponds to an \ic{#pragma omp
  sections}).
\emph{Context nodes} provide additional information on the variables and
parameters, e.g., tensor extents or GPU grid/block sizes; they may also
introduce local scopes and parameters constant within a subtree, which is
useful when mapping induction variables to block and thread identifiers in
CUDA.
Finally, \emph{extension nodes} introduce auxiliary computations that are not
part of the original iteration domain, which is useful for, e.g., introducing
statements copying data to and from shared memory.

  \begin{figure*}[h!tb]
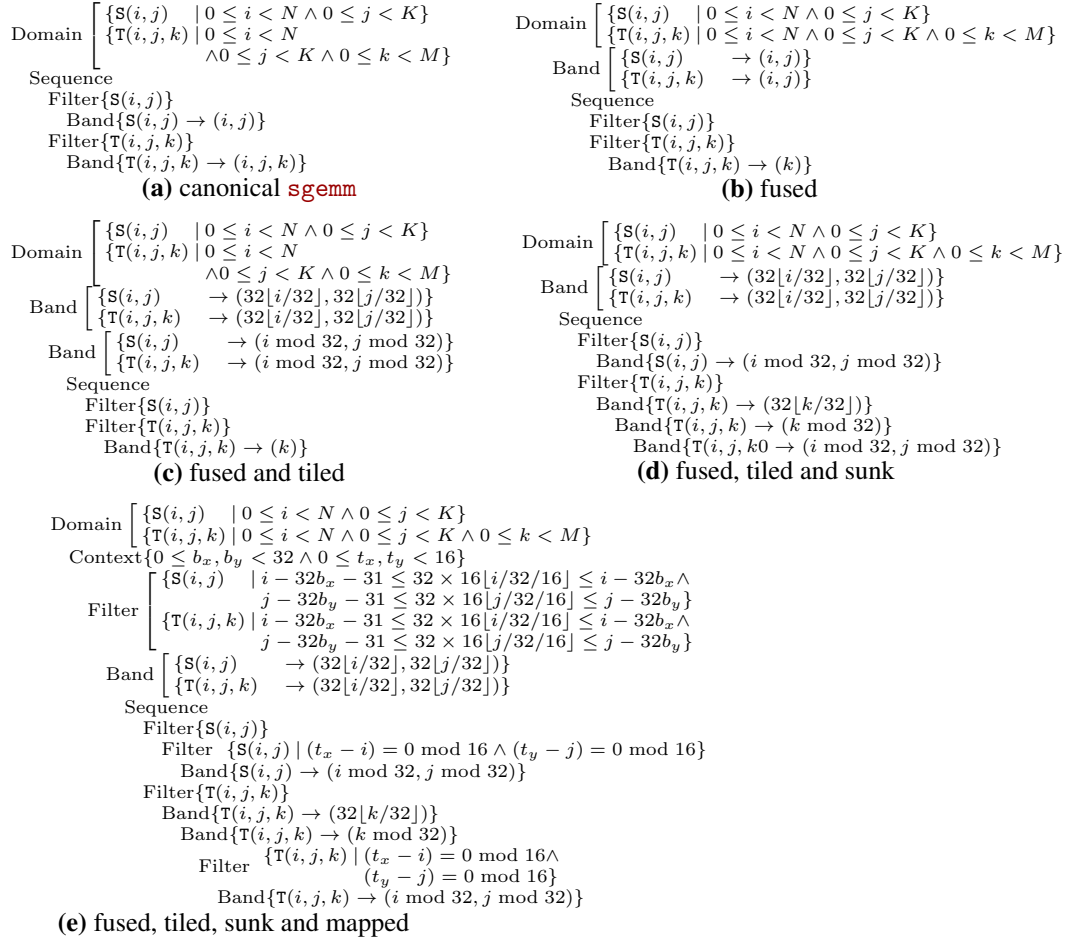

    \begin{minipage}{.48\textwidth}
      {\fontsize{3}{3}\selectfont\scriptsize
        $\displaystyle
          \begin{array}{l}
            \mathrm{Domain}
            \left[\hspace{-0.5em}\begin{array}{l@{~}c@{~}l}
                                   \{ \mathtt{S}(i,j) & \mid & 0 \leq i < N \wedge 0 \leq j < K \} \\
                                   \{ \mathtt{T}(i,j,k) & \mid & 0 \leq i < N \\
                                                      & & \wedge 0 \leq j < K \wedge 0 \leq k < M \} \\
                                 \end{array}\right. \\
            \quad \mathrm{Sequence} \\
            \quad\quad \mathrm{Filter} \{\mathtt{S}(i,j)\} \\
            \quad\quad\quad \mathrm{Band} \{ \mathtt{S}(i,j) \rightarrow (i,j) \} \\
            \quad\quad \mathrm{Filter} \{\mathtt{T}(i,j,k)\} \\
            \quad\quad\quad \mathrm{Band} \{ \mathtt{T}(i,j,k) \rightarrow (i,j,k) \} \\
          \end{array}
          $}
            \par
      \centering
      \textbf{(a)} canonical \ic{sgemm}
    \end{minipage}%
    \begin{minipage}{.51\textwidth}
      {\fontsize{3}{3}\selectfont\scriptsize
        $\displaystyle
          \begin{array}{l}
            \mathrm{Domain}
            \left[\hspace{-0.5em}\begin{array}{l@{~}c@{~}l}
                                   \{ \mathtt{S}(i,j) & \mid & 0 \leq i < N \wedge 0 \leq j < K \} \\
                                   \{ \mathtt{T}(i,j,k) & \mid & 0 \leq i < N \wedge 0 \leq j < K \wedge 0 \leq k < M \} \\
                                 \end{array}\right. \\
            \quad\quad \mathrm{Band}
            \left[\hspace{-0.5em}\begin{array}{ll}
                                   \{ \mathtt{S}(i,j) &\rightarrow (i,j) \} \\
                                   \{ \mathtt{T}(i,j,k) &\rightarrow (i,j) \} \\
                                 \end{array}\right. \\
            \quad\quad\quad \mathrm{Sequence} \\
            \quad\quad\quad\quad \mathrm{Filter} \{\mathtt{S}(i,j)\} \\
            \quad\quad\quad\quad \mathrm{Filter} \{\mathtt{T}(i,j,k)\} \\
            \quad\quad\quad\quad\quad \mathrm{Band} \{ \mathtt{T}(i,j,k) \rightarrow (k) \}
          \end{array}
          $}
            \par
      \centering
      \textbf{(b)} fused
    \end{minipage}

    \medskip
    \begin{minipage}{.48\textwidth}
      {\fontsize{3}{3}\selectfont\scriptsize
        $\displaystyle
          \begin{array}{l}
            \mathrm{Domain}
            \left[\hspace{-0.5em}\begin{array}{l@{~}c@{~}l}
                                   \{ \mathtt{S}(i,j) & \mid & 0 \leq i < N \wedge 0 \leq j < K \} \\
                                   \{ \mathtt{T}(i,j,k) & \mid & 0 \leq i < N \\
                                                      & & \wedge 0 \leq j < K \wedge 0 \leq k < M \}
                                 \end{array}\right.\\
            \quad \mathrm{Band}
            \left[\hspace{-0.5em}\begin{array}{ll}
                                   \{ \mathtt{S}(i,j) &\rightarrow (32 \lfloor i/32 \rfloor, 32 \lfloor j/32 \rfloor) \} \\
                                   \{ \mathtt{T}(i,j,k) &\rightarrow (32 \lfloor i/32 \rfloor, 32 \lfloor j/32 \rfloor) \}
                                 \end{array}\right.\\
            \quad\quad \mathrm{Band}
            \left[\hspace{-0.5em}\begin{array}{ll}
                                   \{ \mathtt{S}(i,j) &\rightarrow (i \bmod 32, j \bmod 32) \} \\
                                   \{ \mathtt{T}(i,j,k) &\rightarrow (i \bmod 32, j \bmod 32) \}
                                 \end{array}\right.\\
            \quad\quad\quad \mathrm{Sequence} \\
            \quad\quad\quad\quad \mathrm{Filter} \{\mathtt{S}(i,j)\} \\
            \quad\quad\quad\quad \mathrm{Filter} \{\mathtt{T}(i,j,k)\} \\
            \quad\quad\quad\quad\quad \mathrm{Band} \{ \mathtt{T}(i,j,k) \rightarrow (k) \}
          \end{array}
          $}
            \par
      \centering
      \textbf{(c)} fused and tiled
    \end{minipage}
    \begin{minipage}{.49\textwidth}
      {\fontsize{3}{3}\selectfont\scriptsize
        $\displaystyle
          \begin{array}{l}
            \mathrm{Domain}
            \left[\hspace{-0.5em}\begin{array}{l@{~}c@{~}l}
                                   \{ \mathtt{S}(i,j) & \mid & 0 \leq i < N \wedge 0 \leq j < K \} \\
                                   \{ \mathtt{T}(i,j,k) & \mid & 0 \leq i < N \wedge 0 \leq j < K \wedge 0 \leq k < M \}
                                 \end{array}\right.\\
            \quad \mathrm{Band}
            \left[\hspace{-0.5em}\begin{array}{ll}
                                   \{ \mathtt{S}(i,j) &\rightarrow (32 \lfloor i/32 \rfloor, 32 \lfloor j/32 \rfloor) \} \\
                                   \{ \mathtt{T}(i,j,k) &\rightarrow (32 \lfloor i/32 \rfloor, 32 \lfloor j/32 \rfloor) \}
                                 \end{array}\right.\\
            \quad\quad \mathrm{Sequence} \\
            \quad\quad\quad \mathrm{Filter} \{\mathtt{S}(i,j)\} \\
            \quad\quad\quad\quad \mathrm{Band} \{\mathtt{S}(i,j) \rightarrow (i \bmod 32, j \bmod 32) \} \\
            \quad\quad\quad \mathrm{Filter} \{\mathtt{T}(i,j,k)\} \\
            \quad\quad\quad\quad \mathrm{Band} \{ \mathtt{T}(i,j,k) \rightarrow (32 \lfloor k/32 \rfloor) \} \\
            \quad\quad\quad\quad\quad \mathrm{Band} \{ \mathtt{T}(i,j,k) \rightarrow (k \bmod 32) \} \\
            \quad\quad\quad\quad\quad\quad \mathrm{Band} \{ \mathtt{T}(i,j,k0 \rightarrow (i \bmod 32, j \bmod 32) \} \\
          \end{array}
          $}
            \par
      \centering
      \textbf{(d)} fused, tiled and sunk
    \end{minipage}

    \medskip
  \begin{minipage}{.45\textwidth}
  {\fontsize{3}{3}\selectfont\scriptsize\hskip15pt
    $\displaystyle\begin{array}{l}
        \mathrm{Domain}
        \left[\hspace{-0.5em}\begin{array}{l@{~}c@{~}l}
                               \{ \mathtt{S}(i,j) & \mid & 0 \leq i < N \wedge 0 \leq j < K \} \\
                               \{ \mathtt{T}(i,j,k) & \mid & 0 \leq i < N \wedge 0 \leq j < K \wedge 0 \leq k < M \}
                             \end{array}\right.\\
        \quad\mathrm{Context} \{ 0 \leq b_x , b_y < 32 \wedge 0 \leq t_x, t_y < 16 \}\\
        \quad\quad \mathrm{Filter}
        \left[\hspace{-0.5em}\begin{array}{l@{~}c@{~}l}
                               \{ \mathtt{S}(i,j) & \mid & i - 32 b_x - 31 \leq 32 \times 16 \lfloor i / 32 / 16 \rfloor \leq i - 32 b_x \wedge \\
                                                  & & j - 32 b_y - 31 \leq 32 \times 16 \lfloor j / 32 / 16 \rfloor \leq j - 32 b_y \} \\
                               \{ \mathtt{T}(i,j,k) & \mid & i - 32 b_x - 31 \leq 32 \times 16 \lfloor i / 32 / 16 \rfloor \leq i - 32 b_x \wedge \\
                                                  & & j - 32 b_y - 31 \leq 32 \times 16 \lfloor j / 32 / 16 \rfloor \leq j - 32 b_y \}
                             \end{array}\right.\\
        \quad\quad\quad \mathrm{Band}
        \left[\hspace{-0.5em}\begin{array}{ll}
                               \{ \mathtt{S}(i,j) &\rightarrow (32 \lfloor i/32 \rfloor, 32 \lfloor j/32 \rfloor) \} \\
                               \{ \mathtt{T}(i,j,k) &\rightarrow (32 \lfloor i/32 \rfloor, 32 \lfloor j/32 \rfloor) \}
                             \end{array}\right.\\
        \quad\quad\quad\quad \mathrm{Sequence} \\
        \quad\quad\quad\quad\quad \mathrm{Filter} \{\mathtt{S}(i,j)\} \\
        \quad\quad\quad\quad\quad\quad \mathrm{Filter}
        \begin{array}{l@{~}c@{~}l}
          \{ \mathtt{S}(i,j) & \mid & (t_x - i) = 0 \bmod 16 \wedge (t_y - j) = 0 \bmod 16 \}
        \end{array}\\
        \quad\quad\quad\quad\quad\quad\quad \mathrm{Band} \{\mathtt{S}(i,j) \rightarrow (i \bmod 32, j \bmod 32) \} \\
        \quad\quad\quad\quad\quad \mathrm{Filter} \{\mathtt{T}(i,j,k)\} \\
        \quad\quad\quad\quad\quad\quad \mathrm{Band} \{ \mathtt{T}(i,j,k) \rightarrow (32 \lfloor k/32 \rfloor) \} \\
        \quad\quad\quad\quad\quad\quad\quad \mathrm{Band} \{ \mathtt{T}(i,j,k) \rightarrow (k \bmod 32) \} \\
        \quad\quad\quad\quad\quad\quad\quad\quad \mathrm{Filter}
        \begin{array}{l@{~}c@{~}l}
          \{ \mathtt{T}(i,j,k) & \mid & (t_x - i) = 0 \bmod 16 \wedge\\
                               & & (t_y - j) = 0 \bmod 16  \}
        \end{array}\\
        \quad\quad\quad\quad\quad\quad\quad\quad\quad \mathrm{Band} \{\mathtt{T}(i,j,k) \rightarrow (i \bmod 32, j \bmod 32) \} \\
    \end{array}$}
        \par
  \centering
  \textbf{(e)} fused, tiled, sunk and mapped
  \end{minipage}

  \caption{Optimization steps for \ic{sgemm} from \figref{tc_sgemm}}
  \label{fig:tree}
\end{figure*}

A \emph{canonical} schedule tree for a TC is defined by an outer Sequence
node, followed by Filter nodes for each TC statement.
Inside each filtered branch, Band nodes define an identity schedule with as
many one-dimensional schedule functions as loop iterators for the statement.
Implicit loops form a permutable band as per TC semantics.
\figref{tree}.a shows the canonical schedule tree for the \ic{sgemm}
TC---declaration of parameters $(N,M,K) \rightarrow \{\dots\}$ is omitted
hereinafter for brevity.

One recognizes a 2-D nest for the initialization statement followed by a 3-D
nest for the update statement.
The schedule can be either parametric in input sizes, or have extra context
information on the tensor sizes.
In cases where Band nodes do not define an injective schedule, the statement
instances are scheduled following the lexicographical order of their domain
coordinates.
Program transformation in the polyhedral model involves defining a different
schedule, which corresponds to a different (partial or total) order of
traversing the iteration domain.
For example, observing that the \texttt{C} tensor is reused between two nests,
one
can construct the schedule in \figref{tree}.b to leverage access locality and
improve performance.
This tree features an outer band node with \ic{i} and \ic{j} loops that became
common to both statements, which corresponds to \emph{loop fusion}.
The sequence node ensures that instances of \ic{S} are executed before
respective instances of \ic{T} enabling proper initialization.
The second band is only applicable to \ic{T} and corresponds to the innermost
(reduction) loop \ic{k}.
Additionally, the tree introduces a Context node to state the assumptions
about the values of parameters.

\paragraph{Out-of-bounds accesses}
The polyhedral model allows for relational encoding of tensor accesses.
Composing those with the iteration domains expressed as sets allows for
computing the set of all accessed tensor elements, i.e., the statement's
footprint, and for checking whether it fits the (specified or inferred) tensor
sizes.
In particular,
access relations enable the detection of out-of-bounds accesses.
Elements that belong to the footprint $\mathcal{F} = \mathcal{D} . \mathcal{A}$,
but not to the set of tensor elements
$\mathcal{T}$, inferred from the tensor shapes, correspond to out-of-bounds
accesses.
Hence, $(\mathcal{D} . \mathcal{A}) \backslash \mathcal{T} = \emptyset$ is a
condition for the absence of out-of-bounds accesses.

\subsection{Polyhedral Transformation and Mapping}
When targeting a parallel architecture, program transformations involve a
change of schedule and also a mapping strategy; these must respect the
dependences while optimizing for target-specific properties.
Beyond guaranteeing the validity of the transformation, dependences can be
used to expose parallelism (independent instances can be executed in parallel)
or to improve data access locality (dependent instances executed close in
time).
Several efficient scheduling algorithms exist, focusing on a combination of
parallelism, locality and
vectorization~\cite{Verdoolaege2017scheduler,Bondhugula2008Pluto,RR-9110,Vasilache2012joint,Pou11}.
Our transformation engine is based on four components:
\begin{normalenumerate}
\item core polyhedral scheduling is provided by \isl, which
  automatically optimizes for (outer) loop parallelism and locality;
  we tuned the affine scheduling heuristic towards folding a complete
  TC program into a single GPU kernel;
\item the schedule is further tiled to facilitate the mapping and
  temporal reuse on the deep parallelism and memory hierarchy of
  GPUs~\cite{Verdoolaege2017scheduler};
\item mapping to GPUs borrows from algorithms previously implemented in
  \ppcg~\cite{PPCG2013}, with additional extensions to support the more
  complex and imperfectly nested structures appearing in ML kernels;
\item memory promotion deals with explicit data transfers to and from
  shared and private memory spaces.  These components and the
  TC-specific extensions are detailed below.
\end{normalenumerate}

\subsection{Scheduling}
The core part of the \isl scheduler iteratively solves integer linear
programming problems to compute piece-wise affine functions that form schedule
bands.
It also ensures that these bands are permutable and can be further tiled.
Internally, it builds a data dependence graph (DDG) where nodes correspond to
statements and edges express dependences between them.
Each edge is annotated with a set of ``typed'' dependence relations.
The \isl scheduler~\cite{Verdoolaege2017scheduler}
is designed for better scalability since integer linear
programming has exponential complexity in the worst case.
In particular, it introduces the \emph{affine clustering} technique that is
based on computing the schedule bands separately for individual
strongly-connected components of the DDG and then clustering these components
iteratively and scheduling with respect to each other.
Clustering not only decreases the size of the linear problems the scheduler
has to solve, but also serves as a basis for \isl's loop fusion heuristic.


We extended the \isl scheduler to provide the caller with a more fine-grained control over the scheduling process.
For affine transformations, the user can supply additional arbitrary constraints to be inserted in the linear program.
For clustering, the user can supply a decision function for pairwise
dependence graph component combination, after it was demonstrated to be valid
by the scheduler.
These configuration points serve as a basis for \emph{scheduling strategies}.
With these strategies, affine transformations can be restricted to:
(1) non-negative schedule coefficients and/or,
(2) non-skewing transformations (i.e., loop interchange and shifting).
Clustering decisions allow for control over the conventional minimum and maximum
fusion targets, and additionally, maximum fusion that preserves at least three
nested parallel loops (to be mapped to CUDA blocks and threads).
Scheduling strategies can be configured and selected through the autotuning
process. In all cases, we enforce that a single GPU kernel is generated.

\subsection{Imperfectly Nested Loop Tiling}
Loop tiling is implemented as a separate step after the scheduling took place
and performed as a schedule tree transformation.
Essentially, it converts a permutable schedule band into a chain of two bands
with the outer band containing tile loops and the inner band containing point
loops with fixed trip count.
This can be seen as a conventional strip-mine and sink transformation.
For example, \figref{tree}.c shows the schedule tree for the fused and tiled \ic{sgemm}.


In addition to conventional loop nest tiling, the schedule tree representation
allows a tiling of imperfectly nested loops.
The technique is based on the following observation: if a loop does not carry
dependences (i.e., is parallel), it can be sunk below any other loop.
In valid schedules, all dependences are carried (or satisfied) by some loop,
along which they feature a positive distance.
A dependence is only violated if it has a negative distance along some loop
\emph{before} it is carried by another loop~\cite{KennedyAllen2002compilers}.
Parallel loops do not carry dependences by definition and therefore do not
affect dependence satisfaction or violation.
Therefore, imperfectly nested tiling is implemented by first tiling all bands
in isolation and then sinking parallel point loops in the tree.
During this process, the point band is replicated in each subtree below a
sequence (or set) node and its schedule is updated so that it only contains
relevant domain points.

The schedule tree for \ic{sgemm} purposefully has two imperfectly nested bands.
Dependence analysis shows that loops \ic{i} and \ic{j} are parallel.
Therefore, we can tile them and sink the point loops below the band of the
reduction \ic{k} loop, resulting in the schedule tree in \figref{tree}.d.
Innermost nested bands with point loops can be joined together into a single
band after checking for permutability.

\subsection{Mapping to Blocks and Threads}
A schedule tree can also be used to represent the \emph{mapping} to an
accelerator, in particular a GPU with multiple blocks and threads.
This operation is performed by associating certain schedule band members, and
the corresponding loops, to thread or block indexes.
Polyhedral code generator then omits the loops, if possible, and rewrites the
index expressions accordingly.
Our mapping algorithm is derived from the one originally designed for \ppcg,
where grid and
block sizes are specified independently from tile sizes.
Due to the semantics of blocks and threads, only parallel loops that belong to a
permutable schedule band can be mapped.
If point loops are mapped to threads, the ratio between tile sizes and blocks
sizes controls the number of iterations executed by each thread.
Note that tile sizes smaller than the block sizes lead to some threads not
performing any computation, if it is the point loops of the tiling that end up
getting mapped to threads.

We require the schedule tree to have at least an outermost band with outer
parallel dimensions.
Contrary to \ppcg, which maps each of possibly multiple outer bands
to both blocks and threads (after tiling), our mapping algorithm maps
a single band to blocks in order to generate a single kernel as
expected by ML frameworks, while mulitple bands can be mapped to threads.
The parallel dimensions of the (single) outermost band are mapped to GPU
blocks.
In each schedule tree branch, the innermost permutable band, typically
consisting of point loops, is mapped to GPU threads with the following
restrictions.
The number of mapped dimensions must be equal across branches.
This may require mapping some thread dimensions to zero in some branches.

The mapping itself is performed by inserting special names, communicated to the
code generator through a context node, and
by associating them to band dimensions in filter nodes.
For the matrix multiplication example, our mapping strategy produces the
schedule tree in \figref{tree}.e.
We introduced a context node in the schedule tree to indicate the effective
sizes of the parameters as well as the grid and block sizes (denoted as $b_x,
b_y$ and $t_x, t_y$, respectively, standing for values potentially taken by
\ic{blockIdx.xy} and \ic{threadIdx.xy}).
This insertion is performed just-in-time, when the effective tensor sizes are
known.
Also pay attention to the Filter nodes referring to $b_x$, $b_y$, $t_x$ and
$t_y$ parameters.
These nodes express the \emph{mapping} to the GPU.

\subsection{Memory Promotion}\label{sec:memory-promotion}
We are interested in promoting parts of tensors into shared or private GPU
memory.
While the promotion decision is taken by a heuristic and the corresponding
imperative code is generated at a later stage, schedule trees offer a
convenient interface for attaching memory-related information.
Memory promotion is based on the notion of \emph{array tile}, a form of data
tiling for software-controlled local memories.
It is a constant-size potentially strided block in the array that covers all
elements accessed within a given (schedule) tile.
We revisit and extend \ppcg's support for memory promotion
\cite{PPCG2013,Verdoolaege2017scheduler}.

\textbf{Promotion of Indirectly Accessed Arrays.} Our memory promotion
approach also handles indirectly accessed arrays.
Without loss of generality, consider the access \ic{O[l+Idx[i][j]][k]}.
We refer to \ic{O} as outer array and to \ic{Idx} as index array.
In case of nested indirections, outer/index pairs are processed iteratively
from innermost to outermost.
While the values taken by the first index expression of the outer array are
unknown statically, we can still cache them locally as
\ic{shared\_O[l][i][j][k]} \ic{= O[l + Idx[i][j]][k]}.
Because some values can be duplicated, indirect promotion is only possible if
both the outer and the index arrays are only read, writing to them could
result in different values that cannot be trivially merged.
In general, we require the index array to have an array tile, i.e., only a
fixed-sized block of it is accessed.
When computing the array tile for the outer array, we ignore the indirect
parts of the subscript (affine parts are treated as usual).
We then introduce as many additional index expressions in the promoted outer
array as there are in the index array.
Extents of the array along these new dimensions correspond exactly to the
array tile sizes of the index array.
Hence an element of the promoted array contains a copy of the global array
element that would be accessed with the given index array.
Indirect subscripts are only used when copying from the global memory while
all other accesses are rewritten in the code generation.
In presence of multiple indirect index expressions that share subexpressions
and have equal tile sizes along the corresponding dimensions, it is sufficient
to introduce a single index expression in the promoted array for all identical
subexpressions.

\textbf{Promotion Heuristics.} Directly accessed arrays are promoted to shared
memory if there exists an array tile of fixed size, if individual elements are
accessed more than once and if at least one of the accesses does not feature
memory coalescing.
The latter is visible from the access relation with the schedule applied to the
domain: the last access dimension should be aligned with the schedule
dimension mapped to \ic{x} threads.
For indirect arrays, the coalescing requirement is dropped because of the
presence of additional long memory dependences that these cases entail.
The total amount of shared memory being fixed, we apply a simple greedy
heuristic and refuse promotion if the required amount of shared memory would
overgrow the available amount.

\subsection{Mapping to Registers}\label{sec:registers}
There are currently no plans to implement more complex register promotion
strategies than those previously supported by \ppcg.
This is a temporary, pragmatic choice, based on the following observations:
\begin{normalitemize}
\item except in limited cases~\cite{SPIRAL,stock2011model}, we have not seen
  empirical evidence that automatically generating low-level code from a
  high-level specification results in a significant performance gain over
  assembly; additionally modern assembly generators are now publicly available
  and re-targetable~\cite{PeachPy};
\item generalizing and re-targeting register optimization passes to different
  architectures with multiple vector lengths, alignment constraints and
  patterns is no easy task;
\item many different strategies exist which include (1) pipelining at the
  register level, (2) register rotation, (3) multi-buffering, (4) new ISAs
  with collective-style semantics~\cite{VoltaMMISA}, some of which are not
  even available without significant genuflections;\footnote{E.g., the mix of
    register banking, $L-1$ FPU operand reuse, and control code on Maxwell
    GPUs, reverse-engineered and exploited in MaxAS
    {(\url{https://github.com/NervanaSystems/maxas})}.}
\item often such strategies are first implemented with intrinsics and assembly
  as a proof of concept and can be easily templated and reused with a tool
  like PeachPy~\cite{PeachPy}.
\end{normalitemize}
Performance results are slightly impacted by this temporary status of the
implementation. We also plan to rely on external library calls, e.g., for
accelerating reductions.

%% file: 5-tuning-and-caching.tex
\section{Autotuning and Caching\label{sec:autotuning}}
While polyhedral scheduling and mapping to GPUs is very cheap compared to
training a neural network, it is far from usable in a traditional JIT
setting. We take advantage of the mostly static structure of neural networks
to cache and reuse the best results of the autotuned compilation pass of an
operation/kernel under similar conditions (input shapes, target architecture,
and optimization options).
We exploit this reuse with a compilation cache.%
\footnote{Note that in a traditional dynamic setting such as stack-RNNs, few
  parameters vary and it is not unreasonable to compile multiple versions; we
  can also relax any dimension to remain parametric (the polyhedral framework
  handles parametric control-flow).}

\subsection{Compilation Cache}
The compilation cache stores the generated CUDA or PTX code for a given
TC. The generated code depends on the input shapes, the
selected optimization options (e.g., tile sizes, loop fusion scheduling
strategy, mapping decisions) and the constraints induced by
the target GPU architecture (e.g., shared memory/register size, fraction of
shared memory to use for occupancy purposes, etc...).
Each cache entry key is therefore a tuple:
$$
\left(\mbox{TC}, \mbox{input}, \mbox{shapes}, \mbox{target}, \mbox{architecture}\right).
$$
For the purpose of caching we use a summarily templatized version of TC to
make the key agnostic to name changes.
Each cache entry holds the fastest known version. Before each kernel
optimization the cache can be queried.
On a cache miss, the regular JIT compilation flow is invoked.
The cache is serialized to a protocol
buffer interchange format~\cite{Protobuf} to enable persistence and reuse.
To avoid long and unpredictable compilation and autotuning times, one may
pre-populate the cache with reference implementations.
Since the cache serializes to strings, we added the
capability to inject entries into the cache
to support particular cases of interest
where low-level manual tuning helps (the experiments in the report do not make
use of this).
In the current implementation, the size of the compilation cache could
potentially become cause for concern: for every operation, the number of input
shapes and optimization options can be arbitrarily large. In practice, however,
the number of operations of interest is limited (e.g., the total number of
different operators in Caffe2 v0.8.0 is approximately 400) and the space of
input shapes is very sparse. Further characterization of these statistics will
be the subject of future work. Importantly, note that our whole pipeline is
parametric and specialization can be injected at any point (or not injected at
all). For the time being we choose to inject runtime sizes as early as
possible in the polyhedral compilation flow (i.e., before polyhedral
scheduling) because it eagerly propagates valuable information for scheduling,
tiling, mapping and total elimination of non-loop control flow (in static
cases).
One good candidate for parametrization is the minibatch dimension. The
minibatch dimension is usually mapped to CUDA grid dimensions. These grid
dimensions are specified by the kernel launch call and can be changed without
the need to generate a different version.
A separate database which includes performance data for all versions is also
maintained for use in autotuning. This will be useful when additional search
techniques are implemented such as Bayesian hyperparameter search
\cite{Snoek2012}, multi-armed bandit optimization \cite{Pacula2012} or
black-box optimization~\cite{Vizier}.

\subsection{Autotuning}
The autotuner interacts with the rest of the environment through the
compilation cache, storing the best versions for later use. It includes the
following steps, setting up:
\begin{normalenumerate}
\item a set of starting configurations that worked well for similar TCs, and a
  few predefined strategies for reference problems;
\item the tuning space dimensions and admissible values for ranges;
\item the type of search---currently a genetic algorithm or random.
\end{normalenumerate}
The autotuner runs for a prescribed amount of time, updating the cache with
better versions along the way. It uses genetic search \cite{Goldberg1989}
which operates on generations of candidate configurations. In each tuning step
the candidates are compiled and profiled, then assigned a fitness value
inversely proportional to their runtime. Each new candidate is bred through
three parent uniform cross-over and also undergoes mutation with a low
probability.
\begin{normalenumerate}
\item three parents are selected probabilistically based on their fitness, the
  higher the fitness the higher the selection chance;
\item each ``gene'', which corresponds to one tuning parameter, of the new
  candidate is randomly selected from the parents.
\end{normalenumerate}
Every new candidate also undergoes mutation: there is a low probability that
zero or more of its genes are assigned random values. This probability is
called the mutation rate and it controls the exploration versus exploitation
tradeoff.

\begin{figure}[h!tb]
\centering\includegraphics[width=.8\columnwidth]{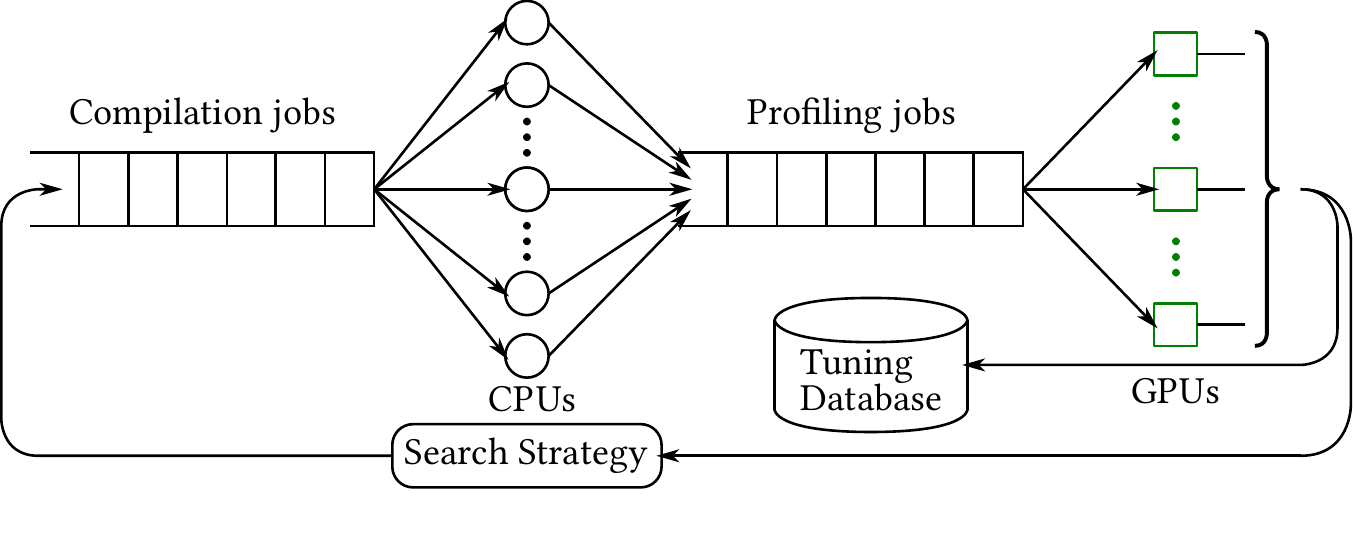}
\caption{Multithreaded autotuning pipeline for kernels}
\label{fig:comp_work_queue}
\end{figure}

Autotuning evaluates 100s to 1000s versions for each kernel.
Search strategies such as genetic algorithms
require evaluating multiple candidates before each search
step; we take advantage of this property by launching multiple compilation
jobs in parallel with our generic multi-threaded, multi-GPU autotuner (see
\figref{comp_work_queue}).
The search strategy enqueues multiple candidate configurations in the
``Compilation Jobs'' queue;
the candidate configurations are compiled in parallel by multiple CPU
threads.
Whenever a compilation job is finished the result is enqueued in the
`Profiling Jobs'' queue; each profiling job is evaluated on an available GPU;
the profiling results are used to update the autotuning database and to
generate the next set of candidates.

A variety of kernel options are tuned:
\begin{normalitemize}
\item the search for tile, block, and grid sizes is narrowed to choices that
  help avoid tail effects; we consider both powers of 2 and integer ceil
  divisors of the problem sizes;\footnote{Indeed, it is common to see problem
    sizes close to a power of $2$ (e.g., $130$) for which it is often better
    to avoid launching an extra block at the cost of a slightly off tile or
    mapping size (e.g., $65$); this brings up to $30\%$ benefits on
    latency-bound kernels.}
\item the bound on how many iterations can be unrolled on the bottom up paths
  from leaves to the root of the schedule tree; we consider powers of $2$ up
  to $4096$;
\item discrete choices on fusion strategies and shapes of the admissible
  schedules, e.g., to prevent loop skewing and trade parallelism for locality;
\item lower-level options such as shared or private memory usage and how
  aggressively to use those memory spaces in each block, which directly
  impacts GPU occupancy.
\end{normalitemize}

%% file: 7-examples.tex
\section{Examples And Performance Results}
\label{sec:examples}

This section reports on the evaluation conducted on an earlier version used
for a submission to the PLDI 2018 conference, relying on a modified version of
the \ppcg\ compiler~\cite{PPCG2013}. The implementation and dependent packages
in the public release of TC differ slightly, bypassing
PENCIL~\cite{Baghdadi2015Pencil} and reimplementing the necessary functions of
\ppcg.

We consider two systems for all experiments:
\begin{normalitemize}
\item 8 Maxwell nodes with 2 socket, 12 core Intel(R) Xeon(R)
CPU E5-2680 v3 @ 2.50GHz, with 8 Tesla M40 GPUs and 12GB of memory each;
\item 8 Pascal nodes with 2 socket, 14 core Intel(R) Xeon(R) CPU
E5-2680 v4 @ 2.40GHz, with 8 Tesla P100-SXM2 GPUs and 16GB of memory each.
\end{normalitemize}
We use CUDA v8.0, CUBLAS v8.0.45, CUDNN v6.0.21, NVRTC
v8.0. Note that with NVRTC v8.0 we experience an extra overhead of around
$15\mu s$ that can become more important than the kernel itself; this is
understood to be reduced in NVRTC v9.0~\cite{VinodGroverPersonal}.
When relevant, we discuss individual kernel performance measured with
\ic{nvprof} but \emph{always report all our numbers
  with the full CPU overhead}, which includes a call to
\ic{cudaDeviceSynchronize}.

We report our baseline and best autotuned variant for every benchmark,
together with the corresponding Caffe2 and ATen reference, when
available.\footnote{ATen---the asynchronous tensor library---currently wraps
  the Torch tensor library and is integrated with PyTorch.  The ATen reference
  implementation is accessible from C++ and acts as a reasonable proxy for the
  expected PyTorch performance; but still not an apples-to-apples comparison.}
Note that Caffe2 is considered as one of the fastest ML frameworks.

The outputs are checked against the reference implementation.
All benchmarks we report are run $1{,}000$ times on the GPU.
We report p0 (a.k.a.\ min), p50 (a.k.a.\ median) and p90 ($90^{\textrm{th}}$ percentile).
The best autotuning results are serialized and
saved in the compilation cache on disk.
A subsequent validation procedure traverses the compilation cache for each
example, selects the best mapping options and runs $1{,}000$ iterations,
checks against the reference implementation and logs the result.
We autotune with 16 CPU threads and 8 GPUs per node. One full sweep of the
genetic autotuner involves a population size of $100$ over $25$ generations
and takes about $6$ hours to complete. We find the bottleneck
is often NVRTC compilation, in particular, we realized it acquires a
global lock internally and can only process one kernel at a time.

Baseline mapping options are picked as ``reasonable guesses'': we did not
spend time performing manual tuning but did not choose a trivially bad solution
either.\footnote{It would be
interesting to compare with Halide's autoscheduler. Unfortunately, it is not
currently functional for GPUs.}
In particular, no iterative, user-involved, transform-inspect-evaluate loop is
required and the polyhedral mapper already automates all decisions and also
offers many different knobs that the autotuner can latch on.
This should come as no surprise given the rich, 30-year history of the
field of polyhedral compilation.
We start our discussion with common kernels that are ubiquitous in ML
workloads. We then show how TC fully automates the synthesis of CUDA kernels
for research layers that do not yet have an existing HPC implementation. We
conclude with a discussion of a model used in production.

Finally, in the context of ever
increasing hardware computation capabilities, models that are
bandwidth-bound today will increasingly shift towards the
latency-bound regime. Specialization of kernel mapping close to that
boundary crossing towards low-latency regimes is obtained
automatically in this contribution.

We provide the baseline and best autotuner options in the
supplementary material, and the final generated code where
informative.

\input{result-table}

\paragraph{Transposed Matrix Multiplication}

In the context of deep learning, transposed matrix multiplication
is ubiquitous:
\begin{tclisting}
def tmm(float(M,K) A, float(N,K) B) -> (C) {
  C(m,n) +=! A(m,kk) * B(n,kk)
}
\end{tclisting}

At the largest problem sizes, our best autotuned version is $4.2\times$
(resp.\ $3.4\times$ on Pascal) \emph{slower} than Caffe2 with CUBLAS.
The reader should note this hints at the improvement
potential of \ourtoolkitname, although:
\begin{normalenumerate}
\item this is one of the most hand-tuned kernel in history and
  CUBLAS operates close to peak at large enough sizes;
\item we do not implement register tiling and advanced promotion schemes for
  now (see Section~\ref{sec:registers}) hence performance is bound by shared
  memory bandwidth;
\item the amount of trickery required to extract the highest levels of
  performance has been documented by Scott Gray~\cite{Maxas}. This involves FU
  operand reuse and SASS code generation to get past register bank conflicts.
  This is not accessible at the CUDA or even PTX levels.
\end{normalenumerate}

On the up side: 
\begin{normalenumerate}
\item our kernel performs close to the peak shared memory bandwidth of the
machine, it is as good as one can hope for in the context of
Section~\ref{sec:registers};
\item it exhibits non-trivial gains in the latency-bound regime;
\item register tiling and reuse schemes are mundane in the polyhedral
  literature and could be implemented in the future;
\item we believe a matching scheme for a library of optimized primitives
  is the most sensible, portable direction,
  as evidenced by Spiral~\cite{SPIRAL},
  Spampinato et al.~\cite{Spampinato16} and the NVIDIA Volta Tensor
  Cores~\cite{VoltaMMISA}; this is work in progress;
\item lastly, our TC abstractions make it very easy to call
  ATen and CUBLAS instead of our generated kernel,
  providing the user with strong
  guarantees performance will not be degraded.
\end{normalenumerate}
Note that on smaller tensors, a.k.a.\ low-latency mode ($28\mu s$ on CPU), on
Maxwell \ic{nvprof} shows our best kernel takes $4.6\mu s$ while CUBLAS
implements a dedicated
kernel ({\ic{maxwell\_sgemm\_128x64\_raggedMn\_tn})
which takes $15.9\mu s$.

\paragraph{Transposed Batched Matrix Multiplication}

This is another
common operation in which batches of matrix pairs are multiplied.
Consider $X$ and $Y$ with dimensions $(B,N,M)$ and $(B,M,K)$
respectively, the corresponding TC is:
\begin{tclisting}
def tbmm(float(B,N,M) X, float(B,K,M) Y) -> (Z) {
  Z(b,n,k) +=! X(b,n,m) * Y(b,k,m)
}
\end{tclisting}
For sizes relevant to Factorization
Machines~\cite{Rendle2010}, $(B,N,M,K)=(500,26,72,26)$,
the \emph{speedup} reaches $3.5\times$ (resp.\ $3.7\times$ on
Pascal) over
CUBLAS---\ic{nvprof} reports $78\mu s$ vs.\ $325\mu s$ for the dedicated
kernel (\ic{maxwell\_sgemmBatched\_128x128\_raggedMn\_nn}).

\paragraph{Grouped Convolutions}

Grouped convolutions have been present for a few years in the DNN research
community, they were already used in the Dropout~\cite{Dropout} network and
have recently gained traction as part of the state-of-the-art
ResNext~\cite{ResNext} model.\footnote{The successor of
  ResNet~\cite{ResNet}, winner of the Imagenet 2015 competition.}
Compared to a traditional 2-D convolution, a grouped convolution takes the
large reduction on the input channel dimension and splits it in two, one
parallel group dimension and one smaller reduction dimension.

A grouped 2D-convolution
with batch size $N$, stride $1$ and no padding is trivially expressible in TC
as:%
\footnote{Incidentally this form is the so-called
  \ic{NCHW} layout, because the $C$ and $F$ dimensions in the TC are referred
  to as channels and filters. Depending on the paper these can also be called
  input and output channels.}
\begin{tclisting}
def gconv(float(N,G,C,H,W) I,float(G,F,C,KH,KW) W1,
          float(M) B) -> (O) {
  O(n,g,o,h,w) +=! I(n,g,i, h + kh, w + kw)
                   * W1(g,o,i,kh,kw)
  O(n,g,o,h,w)  =  O(n,g,o,h,w) + B(m)
}
\end{tclisting}
Despite \ourtoolkitname's current support for time-domain convolutions only,
it still outperforms Caffe2 based on CUDNN
Winograd kernels by $1.4\times$ to $3.6\times$ (resp.\ $1.9\times$ to $8.8\times$ on Pascal). On Maxwell, Caffe2 calls a CUDNN kernel
(\ic{maxwell\_scudnn\_winograd\_128x128\_ldg1\_ldg4\_tile228n\_nt})
$32$ times:
this is about $10\times$ better than a few months earlier when it was
implemented as $32^2$ direct-convolution CUDNN kernel calls;
bridging the years spanned between the discovery of an operation and its
implementation in a library is one of our main contributions.

\paragraph{Production Models}
We finish our discussion with a model used in practice at scale.
Historically, Multi-Layer Perceptrons (MLPs) are the first neural networks in
the literature~\cite{MinskyPapert}.
In the context of
inference---deploying a trained model in production to drive decisions to
improve user experience and revenue, based on individual preferences and
history---the landscape is heavily shifting to custom hardware where the most
serious efficiency gains lie, both in data centers~\cite{TPU17,Catapult} and
on mobile devices.
Still, these models need to be trained before being deployed.
Training usually involves large amounts of time-dependent information such as
weather conditions or world events.
The training procedure generally involves backpropagation and stochastic
gradient descent.
For now, the preferred hardware for this task remains clusters of NVIDIA GPUs.
Because of the small low-latency nature of these models, default library
implementations are not optimal.\footnote{In practice, this limitation has
  become so dire that engineers have expended significant effort writing a
  multi-GPU, asynchronous, Hogwild~\cite{Hogwild} version of the model to
  improve throughput without resorting to fully synchronous minibatch size
  increase.}
One specific MLP model---used in some applications at Facebook---is very
small by the standards of computer vision pipelines~\cite{ResNet}.
Batch size is $128$, mandated by model training accuracy.
Similar models have been proposed in the literature for recommender
systems~\cite{ChengKHSCAACCIA16} and click through rate
prediction~\cite{Zhou2017}.
A TC representation for the neural network portion of the model is given in
the supplementary material.
In the reference Caffe2 implementation, the
\ic{DNN} represents roughly $70\%$ of the time.
It consists of $4$ sections on which we apply \ourtoolkitname: (1) $2$
parallel lookup table operators (\ic{2LUT}) for which we can generate a fused
kernel, (2) $1$ transposed matmul layer, (3) $1$ \ic{MLP1} layer
(a.k.a.\ \ic{fcrelu})
which we find beneficial to distribute from the layers below, and (4)
$3$ \ic{MLP} layers in a sequence (\ic{MLP3}) for which we generate a single
CUDA kernel.
The TC for \ic{2LUT} is:
\begin{tclisting}
def 2LUT(float(E1,D) LUT1, int(B,L1) I1,
         float(E2,D) LUT2, int(B,L2) I2) -> (O1, O2) {
  O1(i,j) +=! LUT1(I1(i,k),j)
  O2(i,j) +=! LUT2(I2(i,k),j)
}
\end{tclisting}

\noindent
\textit{Lookup Table Embeddings.} Large scale
embeddings~\cite{nickel2016review} are used
as a portable encoding.
Concretely, an embedding is a 2-D matrix of numbers.
The associated computation performs a sparse lookup from a
small subset of rows out of a large table and a reduction on the rows.
The sizes of interest are $(E1,E2,D,L1,L2)=(10^7,10^7,64,50,50)$.
We developed a novel 2-stage loading in shared memory (see
Section~\ref{sec:memory-promotion}). Without it, we observe
long latency operations that depend upon each other and cripple the overall
performance by more than $5\times$.
Notice fused \ic{2LUT} performing slightly worse than $2\times$\ic{1LUT}.
An extra order of parallelism is missing without parallel reduction support.
Our synthesized kernel runs $4\times$ (resp.\ $4.1\times$ on Pascal) faster
that the Caffe2 reference, yet the latter relies on the CUB
library and uses this extra parallelism.

\noindent
\textit{Transposed Matmul.} The transposed matrix
multiplication (\ic{C3} is with a weight
matrix of size $(1000,1024)$. The same register and reduction support argument
applies but for such sizes we are already competitive on Pascal.

\noindent
\textit{Single Multi-Layer Perceptron.} The intermediate \ic{MLP1} layer
(i.e., \ic{fcrelu}) is not beneficially fused in frameworks with the following layers. We
achieve up to $1.5\times$ speedup over CUBLAS on Pascal in spite of
missing register optimization.

\noindent
\textit{Fused Multi-Layer Perceptron.} Lastly, \ic{MLP3}
is a sequence of $3$ low-latency \ic{MLP} layers that feed into the binary classifier.
Depending on the ML framework, the underlying HPC library
\emph{and the ML user}, one expects to see anywhere between $3$
and $9$ CUDA kernel calls.
\emph{In this low-latency regime, CUDA kernel launch overhead is significant}.
In contrast, our JIT flow emits $1$ single CUDA kernel call which
runs up to $3.6\times$ faster than the Caffe2 reference, based on a mix of CUDNN
and CUBLAS. Also note our kernel runs in $32\mu s$ with a mean overhead of
$25\mu s$, partly attributable to NVRTC.

%% file: result-table.tex
{
\begin{figure*}
\fontsize{6.5}{6.5}\selectfont
\begin{minipage}[t]{\textwidth}
\textbf{\normalsize Common and Research Kernels}
\par
\input{results_TeslaM40_consolidated_TransposedMatMul}
\input{results_TeslaM40_consolidated_GroupConvolution}
\end{minipage}

\caption{\label{fig:results-research}Wall-clock execution of kernels (in $\mu
  s$). Each kernel ran 1000 times.
  The top half of each table is Tesla~M40
  (Maxwell) and the bottom half is Tesla~P100 (Pascal); N/A indicates the
  framework lacked an implementation}
\end{figure*}
}

{
\begin{figure*}
\fontsize{8}{8}\selectfont
\begin{minipage}[t]{\textwidth}
\textbf{\normalsize \rlap{Production Models}}
\par
\input{results_TeslaM40_consolidated_1LUT}
\input{results_TeslaM40_consolidated_MLP1}
\end{minipage}

\caption{\label{fig:results-prod}Wall-clock execution of kernels (in $\mu s$). Each
  kernel ran 1000 times.
  The top half of each table is Tesla~M40
  (Maxwell) and the bottom half is Tesla~P100 (Pascal); N/A indicates the
  framework lacked an implementation}
\end{figure*}
}

\begin{figure*}[h!tb]
  \begin{minipage}[b]{\textwidth}
    \includegraphics[width=1.06\textwidth]{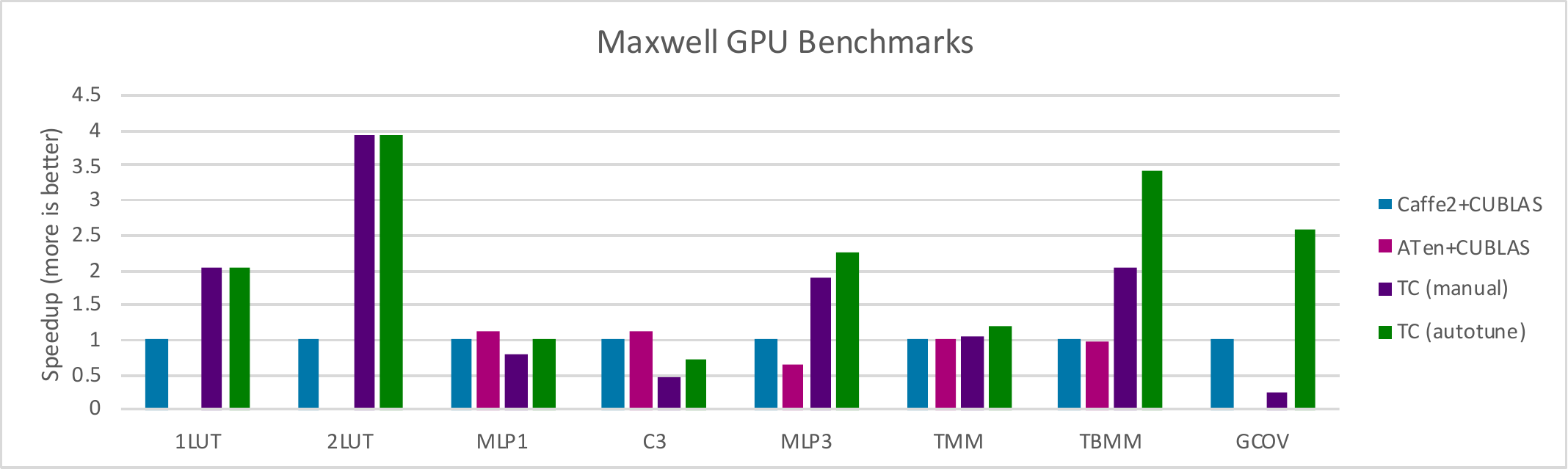}
    \includegraphics[width=1.06\textwidth]{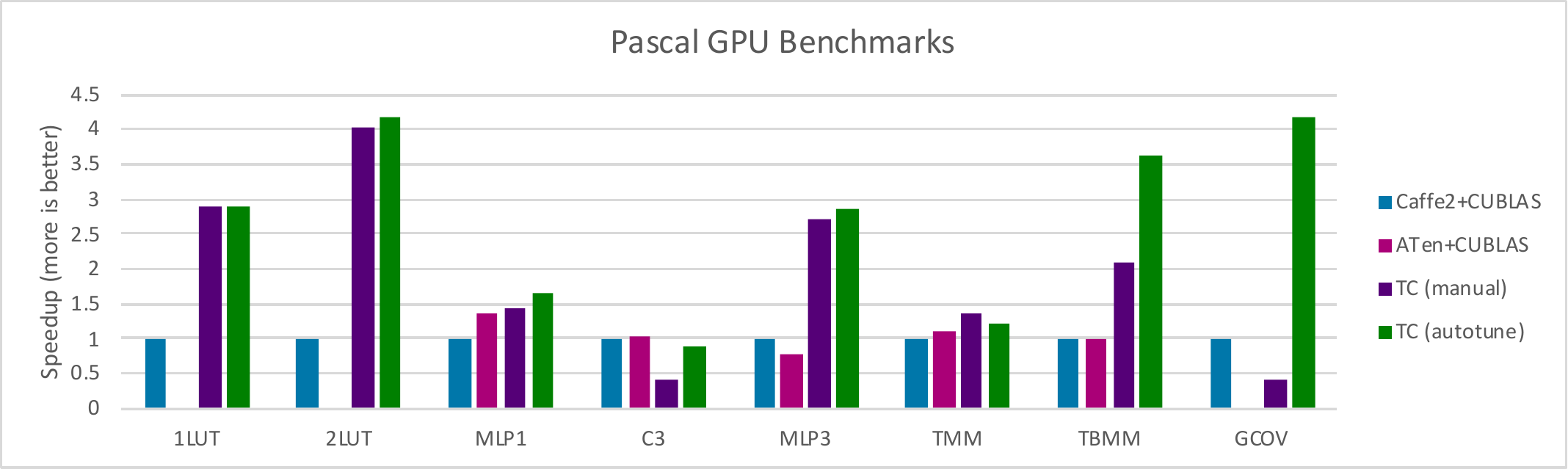}

    \caption{\label{fig:results-graph}Speedup of the \textmd{p50} (median)
      run time across configurations
      and benchmarks, for the smallest data sets,
      normalized to Caffe2-CUBLAS}
  \end{minipage}
\end{figure*}

%% file: results_TeslaM40_consolidated_TransposedMatMul.tex
\begin{tab}{llrrrrrrrrr|lrrr}
\toprule

\multicolumn{2}{c}{$(B,M,K,N)$}  & \multicolumn{3}{c}{$(\text{nil}, 128, 32, 256)$} &\multicolumn{3}{c}{$(\text{nil}, 128, 1024, 1024)$} &\multicolumn{3}{c}{$(\text{nil}, 128, 4096, 16384)$} && \multicolumn{3}{c}{$(500,72,26,26)$}\\

\cmidrule(lr){3-5}
\cmidrule(lr){6-8}
\cmidrule(lr){9-11}
\cmidrule(lr){13-15}

&   &  \text{p0} & \text{p50} & \text{p90}  &  \text{p0} & \text{p50} & \text{p90}  &  \text{p0} & \text{p50} & \text{p90} &&  \text{p0} & \text{p50} & \text{p90} \\

\midrule

\addlinespace[0.5ex]
\multirow{4}{*}{\rotatebox[origin=c]{90}{\textbf{tmm}}}
 & Caffe2 + CUBLAS              & 33 & 35 & 36 &127 & 134 & 136 & 3,527 &  3,578 &  3,666 &
\multirow{4}{*}{\rotatebox[origin=c]{90}{\textbf{tbmm}}}
& 340 & 347 & 350 \\
 &  ATen + CUBLAS               & 35 & 35 & 36 &120 & 123 & 125 & 3,457 &  3,574 &  3,705 &&   342 & 348 & 353 \\

 &  \ourtoolkitname~(manual)    & 32 & 33 & 35 &441 & 446 & 469 &24,452 & 24,583 & 24,656 &&   166 & 170 & 172 \\
 &  \ourtoolkitname~(autotuned) & 28 & 29 & 30 &309 & 313 & 316 &14,701 & 14,750 & 14,768 &&   96 & 101 & 110 \\

\addlinespace[0.5ex]
\hline

\addlinespace[0.5ex]
\multirow{4}{*}{\rotatebox[origin=c]{90}{\textbf{tmm}}}
 & Caffe2 + CUBLAS              & 29 & 30 & 31 &107 & 108 & 109 &2,404 & 2,431 & 3,068 &
\multirow{4}{*}{\rotatebox[origin=c]{90}{\textbf{tbmm}}}
& 189 & 192 & 197 \\

 &  ATen + CUBLAS               & 26 & 27 & 27 &104 & 106 & 108 &2,395 & 2,409 & 3,043 && 188 & 190 & 191 \\
 &  \ourtoolkitname~(manual)    & 21 & 22 & 23 &188 & 194 & 210 &8,378 & 8,402 & 8,411 & &91 & 92 & 93 \\
 &  \ourtoolkitname~(autotuned) & 24 & 25 & 26 &107 & 110 & 111 &8,130 & 8,177 & 8,251 & &51 & 53 & 54 \\

\bottomrule
\end{tab}

%% file: results_TeslaM40_consolidated_GroupConvolution.tex
\begin{tab}{llrrrrrrrrrrrr}
\addlinespace[0.5ex]
\multicolumn{2}{c}{$(N,G,F,C,W,H)$} & \multicolumn{3}{c}{$(32,32,16,16,14,14)$} &\multicolumn{3}{c}{$(32,32,32,32,7,7)$} &\multicolumn{3}{c}{$(32,32,4,4,56,56)$} &\multicolumn{3}{c}{$(32,32,8,8,28,28)$} \\

\cmidrule(lr){3-5}
\cmidrule(lr){6-8}
\cmidrule(lr){9-11}
\cmidrule(lr){12-14}

&   &  \text{p0} & \text{p50} & \text{p90} &  \text{p0} & \text{p50} & \text{p90} &  \text{p0} & \text{p50} & \text{p90} &  \text{p0} & \text{p50} & \text{p90}\\

\midrule

\addlinespace[0.5ex]
 \multirow{4}{*}{\rotatebox[origin=c]{90}{\textbf{gconv}}}
 & Caffe2 + CUDNN          & 1,672 & 1,734 & 1,764 &1,687 & 1,777 & 1,802 &4,078 & 4,179 & 4,206 &3,000 & 3,051 & 3,075 \\
 &  ATen            &  N/A & N/A & N/A &  N/A & N/A & N/A &  N/A & N/A & N/A &  N/A & N/A & N/A \\
 &  \ourtoolkitname~(manual) & 6,690 & 6,752 & 6,805 &3,759 & 3,789 & 3,805 &2,866 & 2,930 & 2,959 &3,939 & 4,009 & 4,045 \\
 &  \ourtoolkitname~(autotuned) & 666 & 670 & 673 &1,212 & 1,215 & 1,216 &1,125 & 1,144 & 1,159 &847 & 863 & 870 \\

\addlinespace[0.5ex]
\hline
\addlinespace[0.5ex]

 \multirow{4}{*}{\rotatebox[origin=c]{90}{\textbf{gconv}}}
 & Caffe2 + CUDNN   &      1,308 & 1,343 & 1,388 &1,316 & 1,338 & 1,350 &4,073 & 4,106 & 4,119 &1,993 & 2,021 & 2,036 \\
 &  ATen            &  N/A & N/A & N/A &  N/A & N/A & N/A &  N/A & N/A & N/A &  N/A & N/A & N/A \\
 &  \ourtoolkitname~(manual) & 3,316 & 3,339 & 3,345 &2,327 & 2,348 & 2,363 &1,683 & 1,691 & 1,694 &1,845 & 1,870 & 1,883 \\
 &  \ourtoolkitname~(autotuned) & 319 & 321 & 322 &691 & 705 & 714 &464 & 481 & 504 &371 & 377 & 379 \\
\bottomrule
\end{tab}

%% file: results_TeslaM40_consolidated_1LUT.tex
\begin{tab}{lrrrrrrrrr}
\toprule

 & \multicolumn{3}{c}{\textbf{1LUT}} &\multicolumn{3}{c}{\textbf{2LUT}} \\

\cmidrule(lr){2-4}
\cmidrule(lr){5-7}

&  \text{p0} & \text{p50} & \text{p90}  &  \text{p0} & \text{p50} & \text{p90} \\

\midrule

\addlinespace[0.5ex]
 Caffe2 + CUBLAS               & 78 & 80 & 82 & 188 & 193 & 207\\
   ATen + CUBLAS               &  N/A & N/A & N/A &  N/A & N/A & N/A\\
   \ourtoolkitname~(manual)    & 38 & 39 & 40 & 47 & 49 & 52 \\
   \ourtoolkitname~(autotuned) & 38 & 39 & 40 & 47 & 49 & 52 \\

\addlinespace[0.5ex]
\hline

\addlinespace[0.5ex]
  Caffe2 + CUBLAS              & 63 & 64 & 66 & 122 & 125 & 128\\

   ATen + CUBLAS               &  N/A & N/A & N/A &  N/A & N/A & N/A\\
   \ourtoolkitname~(manual)    & 21 & 22 & 23 & 30 & 31 & 32\\
   \ourtoolkitname~(autotuned) & 21 & 22 & 23 & 30 & 30 & 31 \\

\bottomrule
\end{tab}

%% file: results_TeslaM40_consolidated_MLP1.tex
\begin{tab}{lrrrrrrrrr}
\addlinespace[0.5ex]

 & \multicolumn{3}{c}{\textbf{MLP1}} &\multicolumn{3}{c}{\textbf{C3}} &\multicolumn{3}{c}{\textbf{MLP3}} \\

\cmidrule(lr){2-4}
\cmidrule(lr){5-7}
\cmidrule(lr){8-10}

&  \text{p0} & \text{p50} & \text{p90}  &  \text{p0} & \text{p50} & \text{p90}  &  \text{p0} & \text{p50} & \text{p90} \\

\midrule

\addlinespace[0.5ex]
 Caffe2 + CUBLAS              & 123 & 125 & 135 & 146 & 159 & 164 & 124 & 128 & 142\\
   ATen + CUBLAS               & 109 & 110 & 112 & 128 & 142 & 148 & 188 & 192 & 213\\
   \ourtoolkitname~(manual)    & 150 & 157 & 159 & 344 & 349 & 351 & 67 & 68 & 70\\
   \ourtoolkitname~(autotuned) & 123 & 125 & 131 & 219 & 224 & 227 & 56 & 57 & 59\\

\addlinespace[0.5ex]
\hline

\addlinespace[0.5ex]
  Caffe2 + CUBLAS              & 123 & 133 & 134 & 107 & 113 & 115 & 129 & 131 & 133\\

   ATen + CUBLAS               & 98 & 98 & 99 & 105 & 110 & 112 & 164 & 167 & 168\\
   \ourtoolkitname~(manual)    & 91 & 92 & 93 & 275 & 279 & 281 & 48 & 48 & 49\\
   \ourtoolkitname~(autotuned) & 79 & 80 & 80 & 117 & 128 & 129 & 45 & 46 & 46\\

\bottomrule
\end{tab}

%% file: 9-conclusion.tex
\section{Perspectives}
This work opens numerous research and exploitation opportunities:
\begin{normalenumerate}
\item distribute and share best implementations, as well as autotuning
  history, for any architecture, via protobuf;
\item port to more architectures, and combine with libraries of primitives
  for high single-thread performance~\cite{Spampinato16};
\item exploit data layout transformations in a systematic manner and use them
  both as part of automated tuning and for easily adding support for vector
  types, in particular on interesting low-precision formats with arbitrary
  number of bits;
\item implement an automated DAG partitioning algorithm, informed by
  underlying performance of synthesized kernels;
\item mix Halide-style transformations for algorithmic tiling,
  model parallelism and model slicing, in combination with advanced loop
  transformations;
\item provide symbolic automatic differentiation on TC directly;
\item more dynamic control flow and gray-box libraries, implemented through
  a mechanism similar to
  PENCIL summaries, as well as dynamic inspection or speculative schemes to
  hide dependences;
\item extend the data representations to sparse, vector and mixed-precision
  types;
\item generally, accelerate ML research with both performance and ease of
  translating a mathematical specification to an actual
  implementation.
\end{normalenumerate}

\section{Conclusion}
We demonstrated an end-to-end flow from the high-level Tensor Comprehensions
(TC) language down to automatically generated kernels on GPUs. TC resembles
the whiteboard mathematical model of a deep neural network and makes it easy
to reason about, communicate, and to manually alter the computations and
storage/computation tradeoffs. The language supports a rich
syntax of tensor operations while remaining within the
realm of polyhedral compiler analysis, affine transformations and (iterated)
linear optimization. The flow leverages decades of progress in polyhedral
compilation and also implements domain-specific optimizations, code
generation, autotuning with a compilation cache, and seamless integration with
Caffe2 and PyTorch through the ATen tensor library.

TC quickly synthesizes solid baseline versions that effectively lift
bottlenecks in large training runs. In practice, when such bottlenecks arise,
ML research slows down sometimes significantly, and serious engineering
efforts need to be mobilized. Our contribution addresses this productivity and
efficiency gap; it brings more expressive power and control in the hands of
domain experts, relieving dependence of ML frameworks on highly tuned vendor
libraries without compromising performance.
TC also automates much boilerplate that has been replicated over the numerous
deep learning frameworks, and builds on a generic polyhedral intermediate
representation and libraries shared with other domains (image processing) and
general-purpose compilers (LLVM/Polly).

\paragraph{Acknowledgements}
We are grateful for the numerous discussions and fruitful ongoing collaboration
with the following people:
Tianqi Chen, Moustapha Ciss\'e, Cijo Jose, Chandan Reddy,
Will Feng, Edward Yang, L\'eon Bottou, Tobias Grosser,
Dmytro Dzhulgokov, and Yangqing Jia.
This work was partly supported by the Swiss National Science
Foundation in the context of grant PZ00P2\_168016.

%% file: a-appendix.tex
\newpage

\appendix

\section{Appendix}
\label{sec:appendix}

This supplementary material collects technical detail useful
to provide complete coverage of the proposed methods.

\label{sec:more_tc}

\figref{tc-grammar} shows the grammar of the Tensor Comprehension language
in EBNF notation.

\subsection{\label{sec:mlframeworks}ML Frameworks of Interest}

We discuss TensorFlow at its core level: mathematical operators, as described originally~\cite{TensorFlow}. Additional higher-level abstractions such as those in Keras~\cite{Keras} and in TensorFlow itself (e.g., tf-fold and tf-eager) improve programmability building upon these operators.
TensorFlow (along with most other ML frameworks) is used to describe and
perform computations on $n$-dimensional tensors. Computation on tensors is
described via \emph{operators}, which take a set of input tensors and define a
set of output tensors as described by a mathematical operation on the input
tensors. Using an initial set of tensors (which may include placeholder or
variable tensors), the programmer composes operators into a static acyclic
dataflow graph called the DAG.
After constructing this DAG, the programmer specifies computations for the
graph to run. Most commonly, one specifies a series of steps of a gradient
descent. The sample code in \figref{tensorflow} gives a feel for using this
framework.

\begin{figure}
\begin{pylisting}
# 1. Construct a graph representing the model.
x = tf.placeholder(tf.float32, [BATCH, 20])
y = tf.placeholder(tf.float32, [BATCH, 10])
w = tf.Variable(tf.random_uniform([20, 10]))
b = tf.Variable(tf.zeros([10]))
z = tf.nn.relu(tf.matmul(x, w) + b)

# 2. Add nodes to define the optimization algorithm.
loss = tf.nn.softmax_cross_entropy_with_logits(z, y)
opt = tf.train.GradientDescentOptimizer(0.1)
train_op = opt.minimize(loss)

# 3. Execute the computation graph on data.
with tf.Session() as sess:
   sess.run(tf.initialize_all_variables())
   for step in range(NUM_STEPS):
       sess.run(train_op, {x: x_data, y: y_data})
\end{pylisting}
\caption{Tensorflow code for training a neural network with one fully-connected layer (adapted from \cite{TensorFlow})}
\label{fig:tensorflow}

\begin{pylisting}
# 1. Construct a graph representing the model.
model = model_helper.ModelHelper(name="train")
x, y = model.net.AddExternalInputs('x','y')
w = model.param_init_net.UniformFill([],'w',shape=[20,10])
b = model.param_init_net.ConstantFill([],'b',shape=[10])
z = model.net.Relu([model.net.Add([\
    model.net.MatMul([x, w], 'z0'), b], 'z1')], 'z')

# 2. Add nodes to define the optimization algorithm.
pred = model.net.Softmax(z, 'pred')
loss = model.net.LabelCrossEntropy([pred, y], 'loss')
model.AddGradientOperators([loss])
optimizer.build_sgd(model, base_learning_rate=0.1, policy="step")

# 3. Execute the computation graph on data.
workspace.RunNetOnce(model.param_init_net)
workspace.CreateNet(model.net)
for i in range(NUM_STEPS):
    workspace.RunNet(model.net.Proto().name)
\end{pylisting}
\caption{Caffe2 code for training a neural network with one fully-connected layer}
\label{fig:caffe2}

\begin{pylisting}
dtype = torch.FloatTensor
x = Variable(x_data, requires_grad=False)
y = Variable(y_data, requires_grad=False)
w = Variable(torch.randn(20, 10).type(dtype), requires_grad=True)
b = Variable(torch.randn(10).type(dtype), requires_grad=True)

learning_rate = 1e-6
for t in range(500):
  y_pred = (x.mm(w)+b).clamp(min=0)
  loss = (y_pred - y).pow(2).sum()
  loss.backward()
  w.data -= learning_rate * w.grad.data
  b.data -= learning_rate * b.grad.data
  
  w.grad.data.zero_()
  b.grad.data.zero_()
\end{pylisting}
\caption{PyTorch code for training a neural network with one fully-connected
layer (from Johnson's sample code \cite{jcjohnsonpytorch})}
\label{fig:pytorch}
\end{figure}

Caffe2~\cite{Caffe2} is structurally close in design to TensorFlow, albeit with a lighter API and making access to the intermediate results in the computation graph much easier. A major design difference is that Caffe2 assumes intermediate objects to be not only tensors but other opaque data structures, such \ic{AFCS} pre-packed matrices of custom layout, or handles to data structures in specialized hardware. This improves portability across different platforms (such as mobile) and interoperability with optimization libraries that may require custom storage formats.
Equivalent syntax for Caffe2 is given in \figref{caffe2}.

PyTorch~\cite{PyTorch} shares many abstraction similarities with TensorFlow and Caffe2, with one major exception: the static computation graph.
Whereas in frameworks such as TensorFlow and Caffe2,
the programmer needs to explicitly specify the computation to be performed and then later execute the computation (potentially given some parameters), PyTorch runs the computation as it is specified.
Equivalent syntax for PyTorch is given in \figref{pytorch}.

MXNet~\cite{MXNet} also shares similar abstractions and offers different paradigms: declarative like Caffe2 and TensorFlow as well as imperative like PyTorch. MXNet also supports multiple languages, Python still being the dominant one. Interestingly, MXNet added compiler abstractions and passes at the core of the toolchain. NNVM is the name of the graph-based intermediate representation (IR) which allows performing graph-rewriting transformations (e.g., for coarsening of library calls and exploring tradeoffs between kernel efficiency and asynchrony like TensorRT~\cite{TensorRT}). NNVM also allows cutting pieces of the graph and delegating those to TVM~\cite{TVM,1802.04799}, a declarative kernel and transformation specification based on Halide~\cite{Halide}.

A number of tensor manipulation ideas underlying modern ML frameworks were first implemented by the Lush/SN3~\cite{lush2002} system in the 1990s. The Lush construct \texttt{idx-bloop} iterates on the first indices of a collection of tensors. Index arithmetic is achieved by preparing the tensor descriptors with helper functions that manipulate the strides associated with each index. For instance, to implement the ``lowering'' of convolutions, the \texttt{unfold} function returns a new tensor descriptor with an additional dimension that scans each of the overlapping convolution windows. The Lush compiler only performs minor optimizations on such loops: tiling and reordering is left to the user.

\subsection{ML Framework Interface API}
\label{sec:interface-api}

We expose tensor comprehensions to ML frameworks with an embedded language API
similar to libraries for regular expressions, SQL queries, or graphics shading
languages. The code is loaded as a string into an execution engine that manages
compilation and execution, as shown in \figref{execution}.
This code can then be run which invokes the JIT compiler and autotuner,
as shown in \figref{run}.
This version of the API is specific to Python and PyTorch, but similar APIs exists for C++ and for other ML frameworks such as Caffe2. 

\begin{figure}[tb]
\begin{minipage}[b]{0.6\columnwidth}
\begin{pylisting}
import tc
ee = tc.ExecutionEngine()
ee.define("""
  def mm(float(M,K) A,
         float(K,N) B) -> (C) {
    C(m,n) +=! A(m,kk) * B(kk,n)
  }
""")
\end{pylisting}
\caption{Build execution engine}
\label{fig:execution}
\end{minipage}
\hfill
\begin{minipage}[b]{0.39\columnwidth}
\begin{pylisting}
import torch
A = torch.randn(3,4)
B = torch.randn(4,5)
C = ee.mm(A, B)
\end{pylisting}
\caption{JIT compile, tune, or hit the compilation cache, then run}
\label{fig:run}
\end{minipage}
\end{figure}

\begin{figure}[tb]
\begin{clisting}
struct TensorInfo {
  DLDataType type;
  std::vector<int64_t> shape;
  std::vector<int64_t> strides;
};

class ExecutionEngine {
 public:
  ExecutionEngine() = default;

  // create the funcDB_ using the language passed to it
  // supports multiple TCs
  void define(const std::string& language);

  // get the output Tensor info that can be used by the 
  // calling framework to allocate storage for the output
  std::vector<TensorInfo> inferOutputTensorInfo(
      const std::string& name,
      const std::vector<const DLTensor*>& inTensorPtrs);

  // Returns a handle for the compiled kernel
  size_t compile(
      const std::string& name,
      const std::vector<const DLTensor*>& inputs,
      const std::vector<DLTensor*>& outputs,
      const IslKernelOptions& options);

  // Run a TC specified by its name on the given tensor 
  // inputs and fill the outputs with the result.
  // The TC is looked up by its handle, sanity checks are 
  // performed on name and input / output sizes.
  // If profile is set, the kernel runtime is returned.
  Duration run(
      const std::string& name,
      const std::vector<const DLTensor*>& inputs,
      const std::vector<DLTensor*>& outputs,
      size_t handle,
      bool profile = false);

 private:
  struct ExecutorInfo {...};

  // Mutex to support multi-threaded autotuner interaction
  std::mutex executorInfoMutex;
  std::vector<std::unique_ptr<ExecutorInfo>> executors_;
  std::map<std::string, tc::TreeRef> tcNameMap_;
  size_t handleCounter = 0;
};
\end{clisting}
\caption{\label{fig:core_api}All libraries that expose tensor comprehensions
go through a framework-agnostic API for loading and running code; to exchange tensor data, the API uses DLTensor, a common interchange format for in-memory tensor data}
\end{figure}





Each of these APIs is just a small wrapper around a core framework-agnostic API shown in Figure~\ref{fig:core_api}.

\begin{figure}[tb]
  {\fontsize{9pt}{9.25pt}%
\begin{tclisting}
num ::= <number literal with C syntax>
id ::= [_a-zA-Z0-9]*[_a-zA-Z][_a-zA-Z0-9]*
exp ::= num
      | ( '-' | '!' | ... ) exp
      | exp ( [+-*/
      | exp '?' exp ':' exp
      | id '.' num # range of num-th dimension of id
      | id '(' exp_list ')' # builtin call or tensor access

reduction ::= <associative reduction operator>
            | '+='  | '*='  | 'min='  | 'max='
            | '+=!' | '*=!' | 'min=!' | 'max=!'

range_constraint ::= id 'in' exp ':' exp

stmt ::= id '(' id_list ')' [ '=' | reduction ] exp
           [ 'where' range_constraint_list ]
       | id_list = id '('id_list ')' # TC function call

arg ::= type id
return ::= id # inferred return type and range

scalar_type ::= 'double' | 'float' | 'half' | 'int' | 'byte' | 'uint32' | ...

type ::= scalar_type [ '(' id_list ')' ]

func ::= # TC function definition
  'def' id '(' arg_list ')' '->' '(' return_list ')' '{'
    stmt_list
  '}'

id_list ::= <comma separated id list>
exp_list ::= <comma separated exp list>
arg_list ::= <comma separated arg list>
stmt_list ::= <whitespace separated stmt list>
return_list ::= <comma separated return list>
range_constraint_list ::= <non-empty comma separated range_constraint list>
\end{tclisting}}
\caption{\label{fig:tc-grammar}EBNF syntax for core TC}
\end{figure}

\subsection{Background on Polyhedral Compilation}
\label{sec:polyhedral_background}

\paragraph{Iteration Domains}
The polyhedral framework operates on individual loop iterations referred to as \emph{statement instances}.
These instances are identified by points in a multidimensional vector space whose coordinates correspond to values of induction variables in the enclosing loops.
Constrained by affine bounds, the set of these points,
referred to as the \emph{iteration domain}, forms a convex polyhedron,
hence the name of the model.
In the context of TC, a statement instance is identified by values of all loop iteration variables that appear in a statement.
The iteration domains are obtained by inferring the ranges of all iterators following the semantics described in Section~\ref{sec:tc}.
Consider again the \ic{sgemm} operation in \figref{tc_sgemm}.
For the sake of brevity, we refer to the initialization statement (line 2) as \ic{S} and to the update statement (line 3) as \ic{T}.
The iteration domain for \ic{sgemm} is then described by the union of sets
\( 
(N, M, K) \rightarrow 
  \{ \mathtt{S}(i,j): 0 \leq i < N \wedge 0 \leq j < K \} \cup 
  \{ \mathtt{T}(i,j,k): 0 \leq i < N \wedge 0 \leq j < K \wedge 0 \leq k < M \}. \)

\paragraph{Schedules}
The points of the \emph{iteration domain} are traversed in the lexicographical order of their logical execution dates.
Conventionally, these dates are defined \textit{via} a
closed-form \emph{schedule} that associates a date to each point. The main
part of the paper sketches the more specific schedule tree representation used in this work.

\paragraph{Code Generation}
Given an iteration domain and a schedule, one can generate a sequence of loop nests that visits all iteration domain points in the order imposed by the schedule.
Efficient algorithms exist to generate \textsc{Fortran}, C + OpenMP, CUDA or OpenCL code~\cite{Bastoul2004Cloog,Chen2012CodeGen+,PPCG2013}.
\isl's state of the art algorithm generates an abstract syntax
tree, which can be traversed to match any imperative syntax~\cite{Grosser2015PolyhedralAST}.

\paragraph{Representing and Preserving Dependences}
One distinctive advantage of the polyhedral framework resides in
its \emph{relational abstraction of dependences} described as unions of
piecewise affine relations---i.e., not necessarily convex or uniform, with the
ability to conduct both exact and conservative/approximate instance-wise
dependence analysis.
This analysis ensures that schedule tree modifications preserve the program semantics.
It is based on \emph{access relations} that associate statement instances to the data they access, in our case---tensor elements.
Access relations use named tuples to identify tensors and are classified according to the access direction: read or write.
For the update statement in the \ic{sgemm} example, the write access relation is
\( \mathcal{A}_\mathrm{write} = (N,M,K) \rightarrow \{ \mathtt{T}(i,j,k) \rightarrow \mathtt{C}(i,j) \} \)
and the read access (union of) relation(s) is
\( \mathcal{A}_\mathrm{read} = (N,M,K) \rightarrow 
\{ \mathtt{T}(i,j,k) \rightarrow \mathtt{A}(i,k) \} \cup
\{ \mathtt{T}(i,j,k) \rightarrow \mathtt{B}(k,j) \} \cup
\{ \mathtt{T}(i,j,k) \rightarrow \mathtt{C}(i,j) \} \cup
\{ \mathtt{T}(i,j,k) \rightarrow \mathtt{a}() \}
\).

\paragraph{Dependence Analysis}
A \emph{dependence} exists between two statement instances that access the same tensor element if at least one of them writes to that element.
For two schedules to give the program identical semantics, it is sufficient that both of them sequence pairs of dependent instances in the same relative order.
Therefore, a program transformation that preserves or \emph{respects} the dependences is considered \emph{valid}.
A dependence relation can be obtained by composing (inverse) access relations and restricting them according to the schedule.
For example, a flow dependence is expressed as 
\( \{ (D_1 \rightarrow D_2) \mid (D_1 \rightarrow D_2) \subset A_{\mathrm{read},D_2} \circ A_{\mathrm{write}, D_1}^{-1} \wedge S(D_1) \prec S(D_2) \}, \)
where $D_1$, $D_2$ are iteration domain spaces and $S$ is the schedule.
In the \ic{sgemm} case, there exists, among others, a flow dependence between the initialization and the update statements, defined by
\(
(N, M, K) \rightarrow \{ \mathtt{S}(i, j) \rightarrow \mathtt{T}(i^\prime, j^\prime, k) \mid 
i^\prime = i \wedge j^\prime = j \wedge
0 \leq i, i^\prime < N \wedge 0 \leq j, j^\prime < K \wedge 0 \leq k < M
\}.
\)
The schedule tree presented in \figref{tree}.a schedules $\mathtt{S}(i,j)$
before any $\mathtt{T}(i,j,k)$ thanks to a Sequence node, respecting the dependence and making the transformation valid.

The previous expression corresponds to the simplest, \emph{memory-based} dependence analysis, which considers two instances to be dependent if they access the same memory location.
It can be further refined by performing exact \emph{data-flow analysis}~\cite{Feautrier1991dataflow}, considering only the instance-wise def-use pairs (a.k.a.\ sources and sinks).
This refinement will result in the flow dependence between \ic{sgemm} statements to be restricted to instances $\mathtt{T}(i,j,k=0)$.
In \isl, dependence analysis also distinguishes
\emph{potential} (may) accesses from \emph{definite} (must) accesses.
The distinction occurs, e.g., when modeling indirect or otherwise non-affine accesses; such accesses are treated as potentially accessing all possible elements along the given tensor dimension.
The data-flow analysis will then include dependences between definite accesses and any potential accesses scheduled between them.
Finally, our dependences can be annotated and further customized after construction.
For example, one can place a ``kill'' statement between two expressions to guarantee (and make that guarantee visible to the polyhedral compiler) that certain (or all) tensor elements are not reused between them.
This is particularly useful when the same (temporary) tensor is reused for storing different values in independent parts of computation, and to bound the liveness of individual tensors across the complete TC graph. This is also important for properly integrating with black-box libraries and conveying semantic information, such as race-free parallelism through internal atomic operations.

Since the TC language does not involve pointer arithmetic and tensor subscripts cannot overflow the inferred sizes, memory aliasing does not occur. This improves the precision of dependence analysis compared to operating on a lower level imperative language \cite{Baghdadi2015Pencil}.

Recall that mapping to blocks and threads implicitly corresponds to tiling, so an array tile can be seen as an over-approximated memory footprint of the block.
Over-approximation enables the promotion of array blocks
of rectangular shape where
axes are aligned with loop iterators, avoiding complex and slow index arithmetic.

\paragraph{Computing Array Tiles}
Memory promotion needs specific array footprint computations and heuristics to group array references accessing overlapping regions (see Section~\ref{sec:memory-promotion}). We recall the state of the art method to achieve this.

A relation describing the array elements accessed inside a block is computed from access relations by applying the tiled and mapped schedule to its domain and then projecting out schedule dimensions that are not mapped to blocks.
In our running matmul example, the write access to \ic{C} becomes
\( \{ [i,j] \rightarrow \mathtt{C}[a_0, a_1] \mid 32 i \leq a_0 \leq 32 i + 31 \wedge 32 j \leq a_1 \leq 32 j + 31 \wedge 0 \leq i,j < N \}. \)
From this relation, we look for a parametric offset such that the image of the relation is constant.
In particular, we consider all lower bounds on index expressions, $a_0$ and $a_1$.
If multiple bounds are possible, the one that results in the smallest size is chosen.
In our example, two bounds are possible $a_0 \geq 0$ and $a_0 \geq 32 i$.
The first one over-approximates the accessed elements by the entire array while the second produces an array tile of $32 \times 32$ elements.
If such offset can be found, the array part is promoted to shared memory.
The promotion itself consists in allocating a buffer in shared memory, copying the data from and to shared memory, and rewriting array accesses in the code.
The copy statements are inserted into the schedule using special Extension nodes immediately before the band mapped to threads.
Our framework attempts to align the last index expressions in copy codes with the \ic{x} thread to maximize coalesced accesses.

\paragraph{Array Reference Grouping and Synchronization}
During promotion, a part of the array is temporarily copied to a buffer; modifying the original array may lead to data inconsistencies.
Therefore, \emph{write} accesses to overlapping parts of the array should be promoted (or not) together.
We perform array reference grouping based on the block-wise access relations.
Array references are grouped if the elements they access in a block overlap and if at least one of them is a write.
If at least one of the grouped references does not have a fixed-size array tile, the entire group cannot be promoted.
Additionally, we also group non-overlapping references if the combined over-approximated array tile would have a smaller size than the sum of sizes of the original array tiles.

Synchronization is inserted after reading the data from global memory, before writing back to global memory and before a write reference and any other overlapping reference to ensure the most recent data from all threads is accessed.

\subsection{Detailed Results on the TC Examples}
\label{sec:appendix-performance}

\input{a-appendix-performance}

%% file: a-appendix-performance.tex


We present the baseline mapping options for all the TC examples
considered and the best autotuned mapping options for non-production
examples.  We show automatically synthesized CUDA where it improves
understanding.  As in Section~\ref{sec:examples}, these results
involved an earlier version of TC for a submission to the PLDI 2018
conference, relying on a modified version of the \ppcg\
compiler~\cite{PPCG2013}.

\subsubsection{Transposed Matrix Multiplication}
We use the following mapping option for all our \ic{tmm} baselines.
\begin{clisting}
tc::IslKernelOptions::makeDefaultMappingOptions()
    .scheduleSpecialize(true)
    .tile({32, 32, 32})
    .mapToThreads({32, 32})
    .mapToBlocks({M / 32, N / 32})
    .useSharedMemory(true)
    .usePrivateMemory(true)
    .unrollCopyShared(true)
    .unroll(256);
\end{clisting}
Given $(M,K,N)=(128,32,256)$ on Maxwell, the autotuner finds:
\begin{clisting}
tc::IslKernelOptions::makeDefaultMappingOptions()
    .scheduleSpecialize(false)
    .tile({4, 32})
    .mapToThreads({1, 32})
    .mapToBlocks({64, 128})
    .useSharedMemory(true)
    .usePrivateMemory(true)
    .unrollCopyShared(false)
    .unroll(4);
\end{clisting}
And produces the following CUDA kernel:
\begin{clisting}
__global__ void tmm_256_32_128(float* __restrict__ C,
  const float* __restrict__ A, const float* __restrict__ B) {
  int b0 = blockIdx.y, b1 = blockIdx.x;
  int t0 = threadIdx.y, t1 = threadIdx.x;
  __shared__ float shared_A[4][33];
  __shared__ float shared_B[32][33];
  float private_C[4][1];

  for (int c2 = 0; c2 <= 3; c2 += 1)
    shared_A[c2][t1] = A[(4 * b0 + c2) * 32 + t1];
  for (int c2 = 0; c2 <= 31; c2 += 1)
    shared_B[c2][t1] = B[(32 * b1 + c2) * 32 + t1];
  __syncthreads();
  for (int c2 = 0; c2 <= 3; c2 += 1) {
    private_C[c2][0] = 0.00000f;
    for (int c4 = 0; c4 <= 31; c4 += 1)
      private_C[c2][0] = (private_C[c2][0] +
        (shared_A[c2][c4] * shared_B[t1][c4]));
  }
  __syncthreads();
  C[4 * b0 * 256 + (32 * b1 + t1)] = private_C[0][0];
  C[(4 * b0 + 1) * 256 + (32 * b1 + t1)] = private_C[1][0];
  C[(4 * b0 + 2) * 256 + (32 * b1 + t1)] = private_C[2][0];
  C[(4 * b0 + 3) * 256 + (32 * b1 + t1)] = private_C[3][0];
  __syncthreads();
\end{clisting}

\subsubsection{Transposed Batched Matrix Multiplication}
We use the following baseline tuning strategy, where $B$ represents the
minibatch size and is mapped to the CUDA processor grid:
\begin{clisting}
tc::IslKernelOptions::makeDefaultMappingOptions()
    .tile({1})
    .mapToThreads({128})
    .mapToBlocks({B})
    .useSharedMemory(true)
    .usePrivateMemory(true)
    .unrollCopyShared(false)
    .unroll(1024);
\end{clisting}
On Maxwell, for $(B,N,M,K)=(500,26,72,26)$, the autotuner finds:
\begin{clisting}
tc::IslKernelOptions::makeDefaultMappingOptions()
    .scheduleSpecialize(true)
    .tile({1})
    .mapToThreads({7, 26})
    .mapToBlocks({72, 16, 1})
    .useSharedMemory(true)
    .usePrivateMemory(true)
    .unrollCopyShared(true)
    .unroll(128);
\end{clisting}
The generated CUDA is unrolled significantly but not enough to remove
indirect expressions in register array variables. We see such patterns that
could be further improved by a more careful mix of tiling, mapping to threads
and unrolling that the autotuner currently does not find. Still the
performance is better than CUBLAS.
\begin{clisting}
// ...
float private_Z[1][4][1];
// ...
for (int c2 = t0; c2 <= 25; c2 += 7) {
  private_Z[0][(-t0 + c2) / 7][0] = 0.000000f;
  private_Z[0][(-t0 + c2) / 7][0] =
    (private_Z[0][(-t0 + c2) / 7][0]
     + (shared_X[0][c2][0] * shared_Y[0][0][t1]));
  // ...
}
\end{clisting}

\subsubsection{Grouped Convolutions}
First of all, one may remark that some incantations do not adopt the
straightforward decomposition: \ic{gconv} and prefer to split the
reduction dimension $\textit{gci}$ into $\textit{gci} \bmod g$ and $\textit{gci} / g$, as shown in the TC below:
\begin{tclisting}
def dconv(float(N,F,H,W) I, float(F,GCI,KH,KW) W1, float(M) B) -> (O) {
  O(n,o,h,w) +=! I(n,gci, h + kh, w + kw) *
                 W1(gci 
  O(n,o,h,w)  =  O(n,o,h,w) + B(m)
}
\end{tclisting}
This semantics is ill-advised. First, it makes indexing dimensions non-affine,
which may be a problem for  dependence analysis. Secondly, it ties a variable
with reduction semantics $i$ to a variable with parallel semantics $g$. As a
consequence, the canonical loop structure coming out of this specification
would require both loops to conservatively degrade to reduction semantics
which is what the group convolutions aim at improving in the first place. In
TC we prefer to write the explicit 5-D version.

We use the following baseline option for all our group convolutions. Note
the specialized threads mapping option to avoid catastrophically bad
performance when $W$ is small.
\begin{clisting}
auto threads = (W >= 10) ?
  std::vector<size_t>{W / 4, H / 2} :
  std::vector<size_t>{4, 8, 4};
auto options = tc::IslKernelOptions::makeDefaultMappingOptions()
    .tile({1, 1, 1})
    .mapToThreads(threads)
    .mapToBlocks({32, 32})
    .useSharedMemory(true)
    .usePrivateMemory(false)
    .unrollCopyShared(true)
    .unroll(1);
\end{clisting}
On Maxwell, for $(W,H)=(7,7)$, the autotuner finds:
\begin{clisting}
auto options = tc::IslKernelOptions::makeDefaultMappingOptions()
    .tile({1, 1})
    .mapToThreads({8, 7, 7})
    .mapToBlocks({32, 32, 3})
    .useSharedMemory(true)
    .usePrivateMemory(false)
    .unrollCopyShared(true)
    .unroll(256);
\end{clisting}
This is not an ideal solution as it does not use registers and it
overprovisions threads to bring in data from global to shared aggressively, as
shown in the code excerpt below,
but it is largely sufficient to deliver much higher performance than a
CUDNN-based implementation.
\begin{clisting}
__global__ gconv...(...) {
  int b0 = blockIdx.y, b1 = blockIdx.x;
  int t0 = threadIdx.z, t1 = threadIdx.y, t2 = threadIdx.x;
  __shared__ float shared_B[1][33];
  __shared__ float shared_I[1][1][32][7][7];
  __shared__ float shared_O[1][1][32][5][5];

  if (t0 == 0 && t1 == 0) {
    shared_B[0][t2] = B[b1 * 32 + t2];
    shared_B[0][t2 + 7] = B[b1 * 32 + (t2 + 7)];
    shared_B[0][t2 + 14] = B[b1 * 32 + (t2 + 14)];
    shared_B[0][t2 + 21] = B[b1 * 32 + (t2 + 21)];
    if (t2 <= 3)
      shared_B[0][t2 + 28] = B[b1 * 32 + (t2 + 28)];
  }
  shared_I[0][0][t0][t1][t2] =
    I[(((b0 * 32 + b1) * 32 + t0) * 7 + t1) * 7 + t2];
  shared_I[0][0][t0 + 8][t1][t2] =
    I[(((b0 * 32 + b1) * 32 + (t0 + 8)) * 7 + t1) * 7 + t2];
  shared_I[0][0][t0 + 16][t1][t2] =
    I[(((b0 * 32 + b1) * 32 + (t0 + 16)) * 7 + t1) * 7 + t2];
  shared_I[0][0][t0 + 24][t1][t2] =
    I[(((b0 * 32 + b1) * 32 + (t0 + 24)) * 7 + t1) * 7 + t2];
  if (t1 <= 4 && t2 <= 4) {
    shared_O[0][0][t0][t1][t2] O
      I[(((b0 * 32 + b1) * 32 + t0) * 5 + t1) * 5 + t2];
    shared_O[0][0][t0 + 8][t1][t2] O
      I[(((b0 * 32 + b1) * 32 + (t0 + 8)) * 5 + t1) * 5 + t2];
    shared_O[0][0][t0 + 16][t1][t2] O
      I[(((b0 * 32 + b1) * 32 + (t0 + 16)) * 5 + t1) * 5 + t2];
    shared_O[0][0][t0 + 24][t1][t2] O
      I[(((b0 * 32 + b1) * 32 + (t0 + 24)) * 5 + t1) * 5 + t2];
  }
  __syncthreads();
  if (t1 <= 4 && t2 <= 4) {
    // ...
    // compute (omitted)
    // ...
  }
  // ...
}
\end{clisting}

\subsubsection{Production Models}
A TC for the portion of interest is given
in~\figref{prodmodel}.\footnote{This is pseudo-code only, TC calling other TCs is not yet supported. In practice we just write independent comprehensions and call them in sequence. The pseudo-code presentation helps understand the global flow of the network while still describing all the details of the computation.}
We do not currently have a generic operator DAG abstraction such as NNVM or
TensorRT. However, the TC
abstraction is malleable enough that we can quickly experiment with
manually grouping
statements in their own TC, which is how we
easily decomposed the model into simpler functions.

\begin{figure}[h!tb]
\begin{tclisting}
def 2LUT(float(E1,D) LUT1, int(B,L1) I1,
         float(E2,D) LUT2, int(B,L2) I2) -> (O1,O2) {
  O1(i,j) +=! LUT1(I1(i,k),j)
  O2(i,j) +=! LUT2(I2(i,k),j)
}
def MLP1(float(B,M) I, float(O,N) W1, float(O) B1) -> (O1) {
  O1(b,n)  = B1(n)
  O1(b,n) += I(b,m) * W1(n,m)
  O1(b,n)  = fmaxf(O1(b,n), 0)
}
def MLP3(float(B,M) I, float(O,N) W2, float(O) B2,
         float(P,O) W3, float(P) B3, float(Q,P) W4,
         float(Q) B4) -> (O1,O2,O3,O4) {
  O2(b,o)  = B2(o)
  O2(b,o) += O1(b,n) * W2(o,n)
  O2(b,o)  = fmaxf(O2(b,o), 0)
  O3(b,p)  = B3(p)
  O3(b,p) += O2(b,o) * W3(p,o)
  O3(b,p)  = fmaxf(O3(b,p), 0)
  O4(b,q)  = B4(q)
  O4(b,q) += O3(b,p) * W4(q,p)
  O4(b,q)  = fmaxf(O4(b,q), 0)
}

def prodModel(float(E1,D) LUT1, int(B,L1) I1,
              float(E2,D) LUT2, int(B,L2) I2,
              float(B,WX) I3, float(WY,WX) W,
              float(N,M) W1, float(N) B1,
              float(O,N) W2, float(O) B2,
              float(P,O) W3, float(P) B3,
              float(Q,P) W4, float(Q) B4)
    -> (C1,C2,C3,I,O1,O2,O3,O4) {
  (C1,C2)    = 2LUT(LUT1,I1,LUT2,I2)
  C3(b,wy)  += I3(b,wxx) * W(wy,wxx)
  I          = concat(C1, C2, C3) # not implemented yet
  O1         = MLP1(I, W1, B1)
  (O2,O3,O4) = MLP3(O1,W2,B2,W3,B3,W4,B4)
  # O4 goes out to binary classifier, omitted here
}
\end{tclisting}
\caption{\label{fig:prodmodel}Full production model (pseudo-code)}
\end{figure}

\paragraph{Lookup Table Embeddings}
We use the following manual option for the $2$ \ic{LUT} kernel and a very
similar one for the single \ic{LUT}:
\begin{clisting}
tc::IslKernelOptions::makeDefaultMappingOptions()
  .tile({4, 32})
  .mapToThreads({1, 32})
  .mapToBlocks({100, 100})
  .useSharedMemory(true)
  .usePrivateMemory(true)
  .unrollCopyShared(true)
  .unrollGpuTile(true)
  .unroll(1024)
\end{clisting}

\paragraph{\ic{C3}} For \ic{C3}, we use the following baseline option:
\begin{clisting}
auto options = tc::IslKernelOptions::makeDefaultMappingOptions()
    .scheduleSpecialize(true)
    .tile({32, 32, 32})
    .mapToThreads({4, 32})
    .mapToBlocks({128, 128})
    .useSharedMemory(true)
    .usePrivateMemory(true)
    .unrollCopyShared(true)
    .unroll(128);
\end{clisting}

\paragraph{\ic{MLP1}} For \ic{MLP1}, we use the following baseline option:
\begin{clisting}
auto options = tc::IslKernelOptions::makeDefaultMappingOptions()
    .scheduleSpecialize(true)
    .tile({16, 16, 128})
    .mapToThreads({16, 16})
    .mapToBlocks({32, 32})
    .useSharedMemory(true)
    .usePrivateMemory(true)
    .unrollCopyShared(false)
    .unroll(1);
\end{clisting}

\paragraph{Fused Multi-Layer Perceptron (\ic{MLP3})}
Recently, ML-oriented libraries such as CUDNN and NNPACK have added
support for trivial pointwise fusion. First, let us note,
that the API for such fused primitives comes with an
impedance mismatch between the library and its concrete integration in ML
frameworks, especially those aiming at targeting multiple hardware platforms.%
\footnote{Brushing
aside the complexities for the library developer for generating a fused
version that effectively outperforms two subsequent matrix-multiplications.}
The collective effort and cognitive overhead spent on such integration is
not trivial. The end user and ML researcher will be impacted
if the framework does not provide a transparent way to access these kernels.
E.g., given a particular
problem size, should one call some \ic{matmul}, \ic{addbias}
and \ic{relu} primitives or the fused \ic{matmul_addbias_relu}? And with
what name or positional parameters? In what order?

For \ic{MLP3}, we use the following baseline option:
\begin{clisting}
IslKernelOptions options = makeDefaultMappingOptions()
    .set_fusion_strategy(FusionStrategy::Max)
    .tile({1})
    .mapToThreads({128})
    .mapToBlocks({128});
\end{clisting}

%% file: tc_extended.bbl
\begin{thebibliography}{10}

\bibitem{TensorFlow}
M.~Abadi, P.~Barham, J.~Chen, Z.~Chen, A.~Davis, J.~Dean, M.~Devin,
  S.~Ghemawat, G.~Irving, M.~Isard, et~al.
\newblock {TensorFlow}: A system for large-scale machine learning.
\newblock In {\em OSDI}, volume~16, pages 265--283, 2016.

\bibitem{Winograd}
S.~G. Andrew~Lavin.
\newblock Fast algorithms for convolutional neural networks.
\newblock {\em CoRR}, abs/1509.09308, 2015, 1509.09308.

\bibitem{Baghdadi2015Pencil}
R.~Baghdadi, U.~Beaugnon, A.~Cohen, T.~Grosser, M.~Kruse, C.~Reddy,
  S.~Verdoolaege, A.~Betts, A.~F. Donaldson, J.~Ketema, J.~Absar, S.~{v
  Haastregt}, A.~Kravets, A.~Lokhmotov, R.~David, and E.~Hajiyev.
\newblock {{PENCIL}}: {{A Platform}}-{{Neutral Compute Intermediate Language}}
  for {{Accelerator Programming}}.
\newblock In {\em 2015 {{International Conference}} on {{Parallel
  Architecture}} and {{Compilation}} ({{PACT}})}, pages 138--149, Oct. 2015.

\bibitem{Clay}
L.~Bagn{\`e}res, O.~Zinenko, S.~Huot, and C.~Bastoul.
\newblock Opening {{Polyhedral Compiler}}'s {{Black Box}}.
\newblock In {\em Proceedings of the 2016 {{International Symposium}} on {{Code
  Generation}} and {{Optimization}}}, CGO 2016, pages 128--138, New York, NY,
  USA, 2016. {ACM}.

\bibitem{PlutoGPU}
M.~M. Baskaran, U.~Bondhugula, S.~Krishnamoorthy, J.~Ramanujam, A.~Rountev, and
  P.~Sadayappan.
\newblock A compiler framework for optimization of affine loop nests for
  {GPGPUs}.
\newblock In {\em Proceedings of the 22Nd Annual International Conference on
  Supercomputing}, ICS '08, pages 225--234, New York, NY, USA, 2008. ACM.

\bibitem{Bastoul2004Cloog}
C.~Bastoul.
\newblock Code {{Generation}} in the {{Polyhedral Model Is Easier Than You
  Think}}.
\newblock In {\em Proceedings of the 13th {{International Conference}} on
  {{Parallel Architectures}} and {{Compilation Techniques}}}, PACT '04, pages
  7--16, Washington, DC, USA, 2004. {IEEE Computer Society}.

\bibitem{VOBLA}
U.~Beaugnon, A.~Kravets, S.~van Haastregt, R.~Baghdadi, D.~Tweed, J.~Absar, and
  A.~Lokhmotov.
\newblock Vobla: A vehicle for optimized basic linear algebra.
\newblock In {\em Proceedings of the 2014 SIGPLAN/SIGBED Conference on
  Languages, Compilers and Tools for Embedded Systems}, LCTES '14, pages
  115--124, New York, NY, USA, 2014. ACM.

\bibitem{Bec03}
O.~Beckmann, A.~Houghton, P.~H.~J. Kelly, and M.~Mellor.
\newblock Run-time code generation in {C++} as a foundation for domain-specific
  optimisation.
\newblock In {\em Proceedings of the 2003 Dagstuhl Workshop on Domain-Specific
  Program Generation}, 2003.

\bibitem{BTO09}
G.~Belter, E.~R. Jessup, I.~Karlin, and J.~G. Siek.
\newblock Automating the generation of composed linear algebra kernels.
\newblock In {\em Proceedings of the Conference on High Performance Computing
  Networking, Storage and Analysis}, SC '09, pages 59:1--59:12, New York, NY,
  USA, 2009. ACM.

\bibitem{Benabderrahmane2010Polyhedral}
M.-W. Benabderrahmane, L.-N. Pouchet, A.~Cohen, and C.~Bastoul.
\newblock The {{Polyhedral Model Is More Widely Applicable Than You Think}}.
\newblock In R.~Gupta, editor, {\em Compiler {{Construction}}}, number 6011 in
  Lecture Notes in Computer Science, pages 283--303. {Springer}, Mar. 2010.

\bibitem{Bondhugula2016Pluto+}
U.~Bondhugula, A.~Acharya, and A.~Cohen.
\newblock The {{Pluto}}+ {{Algorithm}}: {{A Practical Approach}} for
  {{Parallelization}} and {{Locality Optimization}} of {{Affine Loop Nests}}.
\newblock {\em ACM Transactions on Programming Languages and Systems},
  38(3):12:1--12:32, Apr. 2016.

\bibitem{Bondhugula2008Pluto}
U.~Bondhugula, A.~Hartono, J.~Ramanujam, and P.~Sadayappan.
\newblock A {{Practical Automatic Polyhedral Parallelizer}} and {{Locality
  Optimizer}}.
\newblock {\em ACM SIGPLAN Notices}, 43(6):101--113, 2008.

\bibitem{lush2002}
L.~Bottou and Y.~LeCun.
\newblock Lush reference manual.
\newblock \url{http://lush.sf.net/doc.html}, 2002.

\bibitem{Chen2012CodeGen+}
C.~Chen.
\newblock Polyhedra {{Scanning Revisited}}.
\newblock In {\em Proceedings of the 33rd {{ACM SIGPLAN Conference}} on
  {{Programming Language Design}} and {{Implementation}}}, PLDI '12, pages
  499--508, New York, NY, USA, 2012. {ACM}.

\bibitem{CHiLL}
C.~Chen, J.~Chame, and M.~Hall.
\newblock Chill: A framework for composing high-level loop transformations.
\newblock Technical report, Technical Report 08-897, U. of Southern California,
  2008.

\bibitem{TVM}
T.~Chen.
\newblock {{TVM}}.
\newblock \url{http://tvmlang.org/2017/08/17/tvm-release-announcement.html},
  Aug. 2017.

\bibitem{MXNet}
T.~Chen, M.~Li, Y.~Li, M.~Lin, N.~Wang, M.~Wang, T.~Xiao, B.~Xu, C.~Zhang, and
  Z.~Zhang.
\newblock {MXNet}: A flexible and efficient machine learning library for
  heterogeneous distributed systems.
\newblock {\em CoRR}, abs/1512.01274, 2015, 1512.01274.

\bibitem{1802.04799}
T.~Chen, T.~Moreau, Z.~Jiang, H.~Shen, E.~Yan, L.~Wang, Y.~Hu, L.~Ceze,
  C.~Guestrin, and A.~Krishnamurthy.
\newblock {TVM}: End-to-end optimization stack for deep learning, 2018,
  arXiv:1802.04799.

\bibitem{ChengKHSCAACCIA16}
H.~Cheng, L.~Koc, J.~Harmsen, T.~Shaked, T.~Chandra, H.~Aradhye, G.~Anderson,
  G.~Corrado, W.~Chai, M.~Ispir, R.~Anil, Z.~Haque, L.~Hong, V.~Jain, X.~Liu,
  and H.~Shah.
\newblock Wide {\&} deep learning for recommender systems.
\newblock {\em CoRR}, abs/1606.07792, 2016, 1606.07792.

\bibitem{Keras}
F.~Chollet et~al.
\newblock Keras.
\newblock \url{https://github.com/fchollet/keras}, 2015.

\bibitem{DBLP:journals/corr/abs-1003-0358}
D.~C. Ciresan, U.~Meier, L.~M. Gambardella, and J.~Schmidhuber.
\newblock Deep big simple neural nets excel on handwritten digit recognition.
\newblock {\em CoRR}, abs/1003.0358, 2010, 1003.0358.

\bibitem{Coh06}
A.~Cohen, S.~Donadio, M.~J. Garzar\'an, C.~Herrmann, O.~Kiselyov, and D.~Padua.
\newblock In search of a program generator to implement generic transformations
  for high-performance computing.
\newblock {\em Science of Computer Programming}, 62(1):25--46, Sept. 2006.
\newblock Special issue on the First MetaOCaml Workshop 2004.

\bibitem{Coh05}
A.~Cohen, S.~Girbal, D.~Parello, M.~Sigler, O.~Temam, and N.~Vasilache.
\newblock Facilitating the search for compositions of program transformations.
\newblock In {\em ACM Intl. Conf. on Supercomputing (ICS)}, pages 151--160,
  Boston, Massachusetts, June 2005.

\bibitem{Torch7}
R.~Collobert, K.~Kavukcuoglu, and C.~Farabet.
\newblock Implementing neural networks efficiently.
\newblock In G.~Montavon, G.~Orr, and K.-R. Muller, editors, {\em Neural
  Networks: Tricks of the Trade}. Springer, 2012.

\bibitem{Don05}
S.~Donadio, J.~Brodman, T.~Roeder, K.~Yotov, D.~Barthou, A.~Cohen, M.~J.
  Garzar\'an, D.~Padua, and K.~Pingali.
\newblock A language for the compact representation of multiple program
  versions.
\newblock In {\em Languages and Compilers for Parallel Computing (LCPC)}, LNCS,
  Hawthorne, New York, Oct. 2005. Springer.
\newblock 15 pages.

\bibitem{PeachPy}
M.~Dukhan.
\newblock Peachpy: A python framework for developing high-performance assembly
  kernels.
\newblock In {\em Proceedings of the 3rd Workshop on Python for High
  Performance and Scientific Computing}, 2013.

\bibitem{Feautrier1991dataflow}
P.~Feautrier.
\newblock Dataflow {{Analysis}} of {{Array}} and {{Scalar References}}.
\newblock {\em International Journal of Parallel Programming}, 20(1):23--53,
  1991.

\bibitem{feautrier92multi}
P.~Feautrier.
\newblock Some {{Efficient Solutions}} to the {{Affine Scheduling Problem}}.
  {{Part II}}. {{Multidimensional Time}}.
\newblock {\em International Journal of Parallel Programming}, 21(6):389--420,
  1992.

\bibitem{Feautrier2011Polyhedron}
P.~Feautrier and C.~Lengauer.
\newblock Polyhedron {{Model}}.
\newblock In D.~Padua, editor, {\em Encyclopedia of {{Parallel Computing}}},
  pages 1581--1592. {Springer}, 2011.

\bibitem{FRAGUELA2012465}
B.~B. Fraguela, G.~Bikshandi, J.~Guo, M.~J. Garzarán, D.~Padua, and C.~von
  Praun.
\newblock Optimization techniques for efficient hta programs.
\newblock {\em Parallel Computing}, 38(9):465 -- 484, 2012.

\bibitem{FFTW}
M.~Frigo and S.~G. Johnson.
\newblock {{FFTW}}: {A}n adaptive software architecture for the {{FFT}}.
\newblock In {\em Acoustics, Speech and Signal Processing, 1998. Proceedings of
  the 1998 {{IEEE}} International Conference on}, volume~3, pages 1381--1384.
  IEEE, 1998.

\bibitem{URUK}
S.~Girbal, N.~Vasilache, C.~Bastoul, A.~Cohen, D.~Parello, M.~Sigler, and
  O.~Temam.
\newblock Semi-{{Automatic Composition}} of {{Loop Transformations}} for {{Deep
  Parallelism}} and {{Memory Hierarchies}}.
\newblock {\em International Journal of Parallel Programming}, 34(3):261--317,
  July 2006.
\newblock URUK.

\bibitem{Goldberg1989}
D.~E. Goldberg.
\newblock {\em Genetic Algorithms in Search, Optimization and Machine
  Learning}.
\newblock Addison-Wesley Longman Publishing Co., Inc., Boston, MA, USA, 1st
  edition, 1989.

\bibitem{Vizier}
D.~Golovin, B.~Solnik, S.~Moitra, G.~Kochanski, J.~E. Karro, and D.~Sculley,
  editors.
\newblock {\em Google Vizier: A Service for Black-Box Optimization}, 2017.

\bibitem{Protobuf}
Protocol buffers developer guide.
\newblock \url{https://developers.google.com/protocol-buffers/docs/overview},
  2017.

\bibitem{XLA}
{XLA}: Domain-specific compiler for linear algebra to optimizes tensorflow
  computations.
\newblock \url{https://www.tensorflow.org/performance/xla}, 2017.
\newblock commit 0f1b88a.

\bibitem{Caffe2}
P.~Goyal, P.~Doll{\'{a}}r, R.~B. Girshick, P.~Noordhuis, L.~Wesolowski,
  A.~Kyrola, A.~Tulloch, Y.~Jia, and K.~He.
\newblock Accurate, large minibatch {SGD:} training {ImageNet} in 1 hour.
\newblock {\em CoRR}, abs/1706.02677, 2017, 1706.02677.

\bibitem{Maxas}
S.~Gray.
\newblock {Maxas: assembler for Nvidia Maxwell architecture}.
\newblock \url{https://github.com/NervanaSystems/maxas}, 2014.

\bibitem{Grosser2015PolyhedralAST}
T.~Grosser, S.~Verdoolaege, and A.~Cohen.
\newblock Polyhedral {{AST Generation Is More Than Scanning Polyhedra}}.
\newblock {\em ACM Transactions on Programming Languages and Systems},
  37(4):12:1--12:50, July 2015.

\bibitem{VinodGroverPersonal}
V.~Grover.
\newblock personal communication, 2017.

\bibitem{ResNet}
K.~He, X.~Zhang, S.~Ren, and J.~Sun.
\newblock Deep residual learning for image recognition.
\newblock {\em CoRR}, abs/1512.03385, 2015, 1512.03385.

\bibitem{Dropout}
G.~E. Hinton, N.~Srivastava, A.~Krizhevsky, I.~Sutskever, and R.~Salakhutdinov.
\newblock Improving neural networks by preventing co-adaptation of feature
  detectors.
\newblock {\em CoRR}, abs/1207.0580, 2012, 1207.0580.

\bibitem{jcjohnsonpytorch}
J.~Johnson.
\newblock Pytorch examples.
\newblock \url{https://github.com/jcjohnson/pytorch-examples}, 2015.
\newblock commit 0f1b88a.

\bibitem{TPU17}
N.~P. Jouppi, C.~Young, N.~Patil, D.~Patterson, G.~Agrawal, R.~Bajwa, S.~Bates,
  S.~Bhatia, N.~Boden, A.~Borchers, R.~Boyle, P.~Cantin, C.~Chao, C.~Clark,
  J.~Coriell, M.~Daley, M.~Dau, J.~Dean, B.~Gelb, T.~V. Ghaemmaghami,
  R.~Gottipati, W.~Gulland, R.~Hagmann, C.~R. Ho, D.~Hogberg, J.~Hu, R.~Hundt,
  D.~Hurt, J.~Ibarz, A.~Jaffey, A.~Jaworski, A.~Kaplan, H.~Khaitan,
  D.~Killebrew, A.~Koch, N.~Kumar, S.~Lacy, J.~Laudon, J.~Law, D.~Le, C.~Leary,
  Z.~Liu, K.~Lucke, A.~Lundin, G.~MacKean, A.~Maggiore, M.~Mahony, K.~Miller,
  R.~Nagarajan, R.~Narayanaswami, R.~Ni, K.~Nix, T.~Norrie, M.~Omernick,
  N.~Penukonda, A.~Phelps, J.~Ross, M.~Ross, A.~Salek, E.~Samadiani, C.~Severn,
  G.~Sizikov, M.~Snelham, J.~Souter, D.~Steinberg, A.~Swing, M.~Tan,
  G.~Thorson, B.~Tian, H.~Toma, E.~Tuttle, V.~Vasudevan, R.~Walter, W.~Wang,
  E.~Wilcox, and D.~H. Yoon.
\newblock In-datacenter performance analysis of a tensor processing unit.
\newblock In {\em Proceedings of the 44th Annual International Symposium on
  Computer Architecture, {ISCA} 2017, Toronto, ON, Canada, June 24-28, 2017},
  pages 1--12, 2017.

\bibitem{KennedyAllen2002compilers}
K.~Kennedy and J.~R. Allen.
\newblock {\em Optimizing {{Compilers}} for {{Modern Architectures}}: {{A
  Dependence}}-{{Based Approach}}}.
\newblock {Morgan Kaufmann Publishers Inc.}, San Francisco, CA, USA, 2002.

\bibitem{Taco}
F.~Kjolstad, S.~Kamil, S.~Chou, D.~Lugato, and S.~Amarasinghe.
\newblock The tensor algebra compiler.
\newblock {\em Proc. ACM Program. Lang.}, 1(OOPSLA):77:1--77:29, Oct. 2017.

\bibitem{Simit}
F.~Kjolstad, S.~Kamil, J.~Ragan-Kelley, D.~I.~W. Levin, S.~Sueda, D.~Chen,
  E.~Vouga, D.~M. Kaufman, G.~Kanwar, W.~Matusik, and S.~Amarasinghe.
\newblock Simit: A language for physical simulation.
\newblock {\em ACM Trans. Graph.}, 35(2):20:1--20:21, Mar. 2016.

\bibitem{loopy}
A.~Kl{\"{o}}ckner.
\newblock Loo.py: transformation-based code generation for gpus and cpus.
\newblock {\em CoRR}, abs/1405.7470, 2014, 1405.7470.
\newblock Proc. of ARRAY 2014: ACM SIGPLAN Workshop on Libraries, Languages,
  and Compilers for Array Programming.

\bibitem{DBLP:conf/pldi/KongVSFPS13}
M.~Kong, R.~Veras, K.~Stock, F.~Franchetti, L.~Pouchet, and P.~Sadayappan.
\newblock When polyhedral transformations meet {SIMD} code generation.
\newblock In {\em {ACM} {SIGPLAN} Conference on Programming Language Design and
  Implementation, {PLDI} '13, Seattle, WA, USA, June 16-19, 2013}, pages
  127--138, 2013.

\bibitem{Alexnet12}
A.~Krizhevsky, I.~Sutskever, and G.~E. Hinton.
\newblock {ImageNet} classification with deep convolutional neural networks.
\newblock In F.~Pereira, C.~J.~C. Burges, L.~Bottou, and K.~Q. Weinberger,
  editors, {\em Advances in Neural Information Processing Systems 25}, pages
  1097--1105. Curran Associates, Inc., 2012.

\bibitem{Alpha}
H.~Le~Verge, C.~Mauras, and P.~Quinton.
\newblock The {{ALPHA}} language and its use for the design of systolic arrays.
\newblock {\em Journal of VLSI signal processing systems for signal, image and
  video technology}, 3(3):173--182, Sep 1991.

\bibitem{Backprop89}
Y.~LeCun, B.~E. Boser, J.~S. Denker, D.~Henderson, R.~E. Howard, W.~E. Hubbard,
  and L.~D. Jackel.
\newblock Handwritten digit recognition with a back-propagation network.
\newblock In {\em Advances in Neural Information Processing Systems 2, {[NIPS}
  Conference, Denver, Colorado, USA, November 27-30, 1989]}, pages 396--404,
  1989.

\bibitem{OoLaLa}
M.~Luj\'{a}n, T.~L. Freeman, and J.~R. Gurd.
\newblock Oolala: An object oriented analysis and design of numerical linear
  algebra.
\newblock In {\em Proceedings of the 15th ACM SIGPLAN Conference on
  Object-oriented Programming, Systems, Languages, and Applications}, OOPSLA
  '00, pages 229--252, New York, NY, USA, 2000. ACM.

\bibitem{RStream}
B.~Meister, N.~Vasilache, D.~Wohlford, M.~M. Baskaran, A.~Leung, and R.~Lethin.
\newblock {\em R-Stream Compiler}, pages 1756--1765.
\newblock Springer, Boston, MA, 2011.

\bibitem{Brainwave17}
Microsoft unveils project brainwave for real-time ai.
\newblock
  \url{https://www.microsoft.com/en-us/research/blog/microsoft-unveils-project-brainwave},
  Aug. 2017.

\bibitem{MinskyPapert}
M.~L. Minsky and S.~Papert.
\newblock {\em Perceptrons: an introduction to computational geometry; 1st ed.}
\newblock MIT, Cambridge, MA, 1969.

\bibitem{Mullapudi2016HaideAutoscheduler}
R.~T. Mullapudi, A.~Adams, D.~Sharlet, J.~Ragan-Kelley, and K.~Fatahalian.
\newblock Automatically scheduling halide image processing pipelines.
\newblock {\em ACM Transactions on Graphics (TOG)}, 35(4):83, 2016.

\bibitem{Mullapudi:2016:ASH:2897824.2925952}
R.~T. Mullapudi, A.~Adams, D.~Sharlet, J.~Ragan-Kelley, and K.~Fatahalian.
\newblock Automatically scheduling halide image processing pipelines.
\newblock {\em ACM Trans. Graph.}, 35(4):83:1--83:11, July 2016.

\bibitem{Polymage}
R.~T. Mullapudi, V.~Vasista, and U.~Bondhugula.
\newblock Polymage: Automatic optimization for image processing pipelines.
\newblock In {\em Proceedings of the Twentieth International Conference on
  Architectural Support for Programming Languages and Operating Systems},
  ASPLOS '15, pages 429--443, New York, NY, USA, 2015. ACM.

\bibitem{nickel2016review}
M.~Nickel, K.~Murphy, V.~Tresp, and E.~Gabrilovich.
\newblock A review of relational machine learning for knowledge graphs.
\newblock {\em Proceedings of the IEEE}, 104(1):11--33, 2016.

\bibitem{NNH99}
F.~Nielson, H.~Nielson, and C.~Hankin.
\newblock {\em Principles of Program Analysis}.
\newblock Springer, 1999.

\bibitem{TensorRT}
Deploying deep neural networks with {Nvidia TensorRT}.
\newblock
  \url{https://devblogs.nvidia.com/parallelforall/deploying-deep-learning-nvidia-tensorrt},
  Apr. 2017.

\bibitem{VoltaMMISA}
Inside {Volta}: The world’s most advanced data center {GPU}.
\newblock \url{https://devblogs.nvidia.com/parallelforall/inside-volta}, May
  2017.

\bibitem{Pacula2012}
M.~Pacula, J.~Ansel, S.~Amarasinghe, and U.-M. O'Reilly.
\newblock {\em Hyperparameter Tuning in Bandit-Based Adaptive Operator
  Selection}, pages 73--82.
\newblock Springer, Berlin, Heidelberg, 2012.

\bibitem{Pou11}
L.-N. Pouchet, U.~Bondhugula, C.~Bastoul, A.~Cohen, J.~Ramanujam,
  P.~Sadayappan, and N.~Vasilache.
\newblock Loop transformations: Convexity, pruning and optimization.
\newblock In {\em 38th ACM Symp. on Principles of Programming Languages
  (POPL)}, Austin, Texas, Jan. 2011.

\bibitem{PouchetFPGA}
L.-N. Pouchet, P.~Zhang, P.~Sadayappan, and J.~Cong.
\newblock Polyhedral-based data reuse optimization for configurable computing.
\newblock In {\em Proceedings of the ACM/SIGDA International Symposium on Field
  Programmable Gate Arrays}, FPGA '13, pages 29--38, New York, NY, USA, 2013.
  ACM.

\bibitem{RStreamTF}
B.~Pradelle, B.~Meister, M.~Baskaran, J.~Springer, and R.~Lethin.
\newblock Polyhedral optimization of tensorflow computation graphs.
\newblock In {\em 6th Workshop on Extreme-scale Programming Tools (ESPT,
  associated with SC'17)}, Nov. 2017.

\bibitem{Pugh1994Static}
W.~Pugh and D.~Wonnacott.
\newblock Static {{Analysis}} of {{Upper}} and {{Lower Bounds}} on
  {{Dependences}} and {{Parallelism}}.
\newblock {\em ACM Trans. Program. Lang. Syst.}, 16(4):1248--1278, July 1994.

\bibitem{SPIRAL}
M.~P{\"{u}}schel, J.~M.~F. Moura, B.~Singer, J.~Xiong, J.~Johnson, D.~Padua,
  M.~Veloso, and R.~W. Johnson.
\newblock Spiral: A generator for platform-adapted libraries of signal
  processing alogorithms.
\newblock {\em The International Journal of High Performance Computing
  Applications}, 18(1):21--45, 2004.

\bibitem{Catapult}
A.~Putnam, A.~Caulfield, E.~Chung, D.~Chiou, K.~Constantinides, J.~Demme,
  H.~Esmaeilzadeh, J.~Fowers, J.~Gray, M.~Haselman, S.~Hauck, S.~Heil,
  A.~Hormati, J.-Y. Kim, S.~Lanka, E.~Peterson, A.~Smith, J.~Thong, P.~Y. Xiao,
  D.~Burger, J.~Larus, G.~P. Gopal, and S.~Pope.
\newblock A reconfigurable fabric for accelerating large-scale datacenter
  services.
\newblock In {\em Proceeding of the 41st Annual International Symposium on
  Computer Architecuture (ISCA)}, pages 13--24. IEEE Press, June 2014.

\bibitem{PyTorch}
{PyTorch}: Tensors and dynamic neural networks in python with strong {GPU}
  acceleration.
\newblock \url{https://pytorch.org}, 2017.

\bibitem{Halide}
J.~Ragan-Kelley, C.~Barnes, A.~Adams, S.~Paris, F.~Durand, and S.~Amarasinghe.
\newblock Halide: a language and compiler for optimizing parallelism, locality,
  and recomputation in image processing pipelines.
\newblock {\em ACM SIGPLAN Notices}, 48(6):519--530, 2013.

\bibitem{Raina:2009:LDU:1553374.1553486}
R.~Raina, A.~Madhavan, and A.~Y. Ng.
\newblock Large-scale deep unsupervised learning using graphics processors.
\newblock In {\em Proceedings of the 26th Annual International Conference on
  Machine Learning}, ICML '09, pages 873--880, New York, NY, USA, 2009. ACM.

\bibitem{Hogwild}
B.~Recht, C.~Re, S.~Wright, and F.~Niu.
\newblock Hogwild: A lock-free approach to parallelizing stochastic gradient
  descent.
\newblock In J.~Shawe-Taylor, R.~S. Zemel, P.~L. Bartlett, F.~Pereira, and
  K.~Q. Weinberger, editors, {\em Advances in Neural Information Processing
  Systems 24}, pages 693--701. Curran Associates, Inc., 2011.

\bibitem{Rendle2010}
S.~Rendle.
\newblock Factorization machines.
\newblock In {\em Proceedings of the 2010 IEEE International Conference on Data
  Mining}, ICDM '10, pages 995--1000, Washington, DC, USA, 2010. IEEE Computer
  Society.

\bibitem{Rompf:2010:LMS:1868294.1868314}
T.~Rompf and M.~Odersky.
\newblock Lightweight modular staging: A pragmatic approach to runtime code
  generation and compiled dsls.
\newblock In {\em Proceedings of the Ninth International Conference on
  Generative Programming and Component Engineering}, GPCE '10, pages 127--136,
  New York, NY, USA, 2010. ACM.

\bibitem{Smi00}
M.~D. Smith.
\newblock Overcoming the challenges to feedback-directed optimization (keynote
  talk).
\newblock In {\em Proceedings of the ACM SIGPLAN Workshop on Dynamic and
  Adaptive Compilation and Optimization}, DYNAMO '00, pages 1--11, New York,
  NY, USA, 2000. ACM.

\bibitem{Snoek2012}
J.~Snoek, H.~Larochelle, and R.~P. Adams.
\newblock Practical bayesian optimization of machine learning algorithms.
\newblock In {\em Proceedings of the 25th International Conference on Neural
  Information Processing Systems - Volume 2}, NIPS'12, pages 2951--2959, USA,
  2012. Curran Associates Inc.

\bibitem{Spampinato16}
D.~G. Spampinato and M.~P{\"{u}}schel.
\newblock A basic linear algebra compiler for structured matrices.
\newblock In {\em Proceedings of the 2016 International Symposium on Code
  Generation and Optimization, {CGO} 2016, Barcelona, Spain, March 12-18,
  2016}, pages 117--127, 2016.

\bibitem{stock2011model}
K.~Stock, T.~Henretty, I.~Murugandi, P.~Sadayappan, and R.~Harrison.
\newblock Model-driven {{SIMD}} code generation for a multi-resolution tensor
  kernel.
\newblock In {\em 2011 IEEE International Parallel \& Distributed Processing
  Symposium (IPDPS)}, pages 1058--1067. IEEE, 2011.

\bibitem{Theano}
{Theano Development Team}.
\newblock {Theano: A {Python} framework for fast computation of mathematical
  expressions}.
\newblock {\em arXiv e-prints}, abs/1605.02688, May 2016.

\bibitem{Latte}
L.~Truong, R.~Barik, E.~Totoni, H.~Liu, C.~Markley, A.~Fox, and T.~Shpeisman.
\newblock Latte: A language, compiler, and runtime for elegant and efficient
  deep neural networks.
\newblock In {\em Proceedings of the 37th ACM SIGPLAN Conference on Programming
  Language Design and Implementation}, PLDI '16, pages 209--223, New York, NY,
  USA, 2016. ACM.

\bibitem{FBFFT15}
N.~Vasilache, J.~Johnson, M.~Mathieu, S.~Chintala, S.~Piantino, and Y.~LeCun.
\newblock Fast convolutional nets with fbfft: A {GPU} performance evaluation.
\newblock {\em CoRR}, abs/1412.7580, 2014.

\bibitem{Vasilache2012joint}
N.~Vasilache, B.~Meister, M.~Baskaran, and R.~Lethin.
\newblock Joint scheduling and layout optimization to enable multi-level
  vectorization.
\newblock In {\em IMPACT-2: 2nd International Workshop on Polyhedral
  Compilation Techniques}, Paris, France, Jan 2012.

\bibitem{VG98}
T.~Veldhuizen and E.~Gannon.
\newblock {{Active libraries}}: {{Rethinking}} the roles of compilers and
  libraries.
\newblock In M.~E. Henderson, C.~R. Anderson, and S.~L. Lyons, editors, {\em
  Proceedings of the 1998 SIAM Workshop: Object Oriented Methods for
  Interoperable Scientific and Engineering Computing}, pages 286--295. SIAM
  Press, 1998.

\bibitem{ISL10}
S.~Verdoolaege.
\newblock Isl: An integer set library for the polyhedral model.
\newblock In {\em Proceedings of the Third International Congress Conference on
  Mathematical Software}, ICMS'10, pages 299--302, Berlin, Heidelberg, 2010.
  Springer.

\bibitem{Verdoolaege2011iscc}
S.~Verdoolaege.
\newblock Counting {{Affine Calculator}} and {{Applications}}.
\newblock In {\em First {{International Workshop}} on {{Polyhedral Compilation
  Techniques}} ({{IMPACT}}'11)}, Chamonix, France, Apr. 2011.

\bibitem{PPCG2013}
S.~Verdoolaege, J.~Carlos~Juega, A.~Cohen, J.~Ignacio~G{\'o}mez, C.~Tenllado,
  and F.~Catthoor.
\newblock Polyhedral {{Parallel Code Generation}} for {{CUDA}}.
\newblock {\em ACM Transactions on Architecture and Code Optimization},
  9(4):54:1--54:23, Jan. 2013.

\bibitem{PET}
S.~Verdoolaege and T.~Grosser.
\newblock Polyhedral extraction tool.
\newblock In {\em In Second International Workshop on Polyhedral Compilation
  Techniques (IMPACT’12)}, 2012.

\bibitem{Verdoolaege2014ScheduleTrees}
S.~Verdoolaege, S.~Guelton, T.~Grosser, and A.~Cohen.
\newblock Schedule {{Trees}}.
\newblock In {\em 4th {{Workshop}} on {{Polyhedral Compilation Techniques}}
  ({{IMPACT}}, {{Associated}} with {{HiPEAC}})}, page~9, Vienna, Austria, Jan.
  2014.

\bibitem{Verdoolaege2017scheduler}
S.~Verdoolaege and G.~Janssens.
\newblock Scheduling for {PPCG}.
\newblock Report CW 706, Department of Computer Science, KU Leuven, Leuven,
  Belgium, June 2017.

\bibitem{ATLAS}
R.~C. Whaley and J.~J. Dongarra.
\newblock Automatically tuned linear algebra software.
\newblock In {\em Proceedings of the 1998 ACM/IEEE Conference on
  Supercomputing}, SC '98, pages 1--27, Washington, DC, USA, 1998. IEEE
  Computer Society.

\bibitem{ResNext}
S.~Xie, R.~B. Girshick, P.~Doll{\'{a}}r, Z.~Tu, and K.~He.
\newblock Aggregated residual transformations for deep neural networks.
\newblock {\em CoRR}, abs/1611.05431, 2016, 1611.05431.

\bibitem{Zhou2017}
G.~{Zhou}, C.~{Song}, X.~{Zhu}, X.~{Ma}, Y.~{Yan}, X.~{Dai}, H.~{Zhu},
  J.~{Jin}, H.~{Li}, and K.~{Gai}.
\newblock {Deep Interest Network for Click-Through Rate Prediction}.
\newblock {\em ArXiv e-prints}, June 2017, 1706.06978.

\bibitem{RR-9110}
O.~Zinenko, S.~Verdoolaege, C.~Reddy, J.~Shirako, T.~Grosser, V.~Sarkar, and
  A.~Cohen.
\newblock {Unified Polyhedral Modeling of Temporal and Spatial Locality}.
\newblock Technical Report RR-9110, Inria, Paris, France, Oct 2017.

\end{thebibliography}
